\documentclass[manuscript, authorversion, screen]{acmart}

\AtBeginDocument{%
  \providecommand\BibTeX{{%
    Bib\TeX}}}

\usepackage[]{natbib}
\usepackage[linesnumbered,ruled]{algorithm2e}

\usepackage{balance}

\usepackage{graphicx}
\usepackage{textcomp}
\usepackage{xcolor}
\usepackage{comment}
\usepackage{booktabs}
\usepackage{multirow}
\usepackage{xspace}
\usepackage{tcolorbox}

\usepackage{array}

\newcommand{\PreserveBackslash}[1]{\let\temp=\\#1\let\\=\temp}
\newcolumntype{C}[1]{>{\PreserveBackslash\centering}p{#1}}
\newcolumntype{R}[1]{>{\PreserveBackslash\raggedleft}p{#1}}
\newcolumntype{L}[1]{>{\PreserveBackslash\raggedright}p{#1}}

\newcommand{\todoc}[2]{{\textcolor{#1}{\textbf{#2}}}}

\newcommand{\jinqiu}[1]{\todoc{purple}{{\it [Jinqiu says: #1]}}}
\newcommand{\triet}[1]{\todoc{blue}{{\it [minhtriet: #1]}}}

\def\BibTeX{{\rm B\kern-.05em{\sc i\kern-.025em b}\kern-.08em
    T\kern-.1667em\lower.7ex\hbox{E}\kern-.125emX}}
    
\newcommand{\pa}[1]{\noindent\textbf{#1}}
\newcommand{\toolS}{SimsV\space}
\newcommand{\tool}{SimsV\xspace}

\newcommand{\svl}{LGSVL}
\newcommand{\svlS}{LGSVL\xspace}
\newcommand{\lidarS}{LiDAR\space}

\newcommand{\RQOne}{Can \toolS identify weakness in the perception module of Apollo?}
\newcommand{\RQThree}{What are the adverse consequences of the perception weakness detected by \tool?}
\newcommand{\rqboxc}[1]{\begin{tcolorbox}[left=4pt,right=4pt,top=4pt,bottom=4pt,colback=gray!5,colframe=gray!40!black,before skip=4pt,after skip=4pt]#1\end{tcolorbox}}

\setcopyright{acmlicensed}
\copyrightyear{2021}
\acmYear{2021}
\acmDOI{XXXXXXX.XXXXXXX}

\acmConference[]{}{2021}{Woodstock, NY}

\begin{document}

\title{Perception-Guided Fuzzing for Simulated Scenario-Based Testing of Autonomous Driving Systems}

\author{Tri Minh Triet Pham}
\affiliation{%
  \institution{Concordia University}
  \city{Montreal}
  \country{Canada}}
\email{p\_triet@encs.concordia.ca}
\author{Bo Yang}
\affiliation{%
  \institution{Concordia University}
  \city{Montreal}
  \country{Canada}}
\email{b\_yang20@encs.concordia.ca}
\author{Jinqiu Yang}
\affiliation{%
  \institution{Concordia University}
  \city{Montreal}
  \country{Canada}}
\email{jinqiu.yang@concordia.ca}

\renewcommand{\shortauthors}{Pham et al.}

\begin{abstract}
% Autonomous Driving Systems (ADS) have gained much interest and their development have made huge progress. 
Autonomous Driving Systems (ADS) have made huge progress and started on-road testing or even commercializing trials. 
ADS are complex and difficult to test: they receive input data from multiple sensors and make decisions using a combination of multiple deep neural network models and code logic.
The safety of ADS is of utmost importance as their misbehavior can result in costly catastrophes, including the loss of human life.
%This has drawn the attention of testing communities to identify and reduce erroneous behaviors. 
%While previous studies have either tested or attacked ADS in various scopes and ways, the analysis often stops at detecting issues. 
In this work, we propose \tool, which performs system-level testing on multi-module ADS. \toolS targets perception failures of ADS and further assesses the impact of perception failure on the system as a whole.
%The Perception Module in ADS serves as the eye of the ADS, which has also been studied several times before. Yet, the testing is incomplete. In this work, we test the ADS by exploiting the weakness in perception. %ind erroneous behavior, with the additional capability to identify if the faulty behavior was caused by the ADS Perception module.
\tool leverages a high-fidelity simulator for test input and oracle generation by continuously applying predefined mutation operators. 
In addition, \toolS leverages various metrics to guide the testing process. 
We implemented a prototype \toolS for testing a commercial-grade Level 4 ADS (i.e., Apollo) using a popular open-source driving platform simulator. Our evaluation shows that \toolS is capable of finding weaknesses in the perception of Apollo. Furthermore, we show that by exploiting such weakness, \toolS finds severe problems in Apollo, including collisions.
%We perform our test by mutation fuzzing of scenarios, maximizing various metrics to generate scenes that can potentially cause faulty behaviors. Then we attempt to maximize the collision chance and identify whether it was caused by Perception Module malfunction. We can successfully detect bugs that were caused by the malfunction of the Perception Modules, for which we will discuss the cause and possible fixes.
\end{abstract}

\begin{CCSXML}
<ccs2012>
   <concept>
       <concept_id>10011007.10011074.10011099.10011102.10011103</concept_id>
       <concept_desc>Software and its engineering~Software testing and debugging</concept_desc>
       <concept_significance>500</concept_significance>
       </concept>
   <concept>
       <concept_id>10010147.10010341.10010349.10010359</concept_id>
       <concept_desc>Computing methodologies~Real-time simulation</concept_desc>
       <concept_significance>500</concept_significance>
       </concept>
   <concept>
       <concept_id>10010147.10010178</concept_id>
       <concept_desc>Computing methodologies~Artificial intelligence</concept_desc>
       <concept_significance>500</concept_significance>
       </concept>
 </ccs2012>
\end{CCSXML}

\ccsdesc[500]{Software and its engineering~Software testing and debugging}
\ccsdesc[500]{Computing methodologies~Real-time simulation}
\ccsdesc[500]{Computing methodologies~Artificial intelligence}

\keywords{autonomous driving systems, software testing, quality assurance}

\maketitle

\section{Introduction}

Due to recent advances in AI  technologies, the software industry is developing various AI-powered systems and applications, such as image recognition, natural language translation, and health care management. In particular, Autonomous driving systems (ADS) have received significant attention from both industry and academia for their great potential and future values.  
%Hence, many technology companies are racing to build the first commercial Level 4 ADS. 
As ADS are run on roads and appear more in the public eye, their safety becomes paramount. Unfortunately, even though ADS have been tested on public roads for tens of thousands of hours~\cite{waymo}, there are still some untested corner cases that result in catastrophic consequences~\cite{news1, news2, news3}. 

%In addition to closed source ADS being developed by automotive manufacturers such as Tesla's Autopilot, there are open source alternatives such as Baidu's Apollo~\cite{apollo}, Autoware Foundations' Autoware~\cite{autoware}, and comma.ai's openpilot~\cite{openpilot}. In particular, Apollo is highly sophisticated and is already operating taxi services in multiple cities around the world~\cite{apollo-go}. 

%different types
%Autonomous Driving Systems (ADS) are being rapidly developed. Notable open-source projects include Baidu's Apollo \cite{apollo}, Autoware Foundations' Autoware \cite{autoware}, and comma.ai's openpilot\cite{openpilot}, where Apollo and Autoware are highly sophisticated and have much potential to reach fully automated status. 

%As ADS are being rapidly developed and their use cases expanded, it has come to the public's attention that this useful technology is difficult to develop; and as useful and convenient as they are, this technology would also have the ability to cause significant damages and death\cite{news1, news2, news3}.

\begin{figure}
\includegraphics[scale=0.8,keepaspectratio]{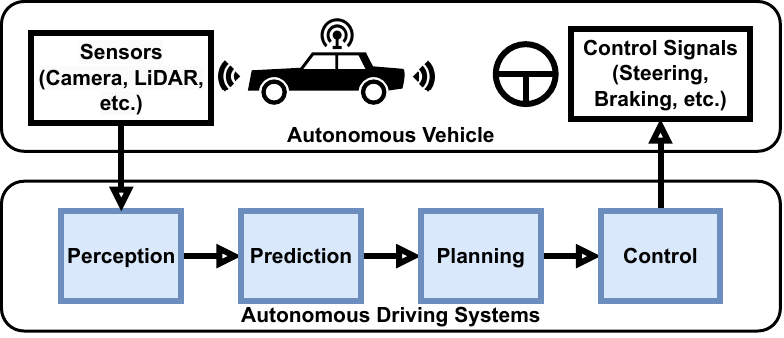}
\caption{An overview of a multi-module ADS.}
\label{fig:fig_ads_generic}
\end{figure}

To this end, many research efforts focus on ensuring the correctness and safety of ADS (i.e., over 80 papers as shown in a recent survey~\cite{surveypaper}). Prior efforts on testing ADS have two main types of ADS under test, i.e., driving models or multi-module ADS. A driving model uses a single deep learning model to process sensor input and to make driving decisions (e.g., steer angle). 
In contrast, a multi-module ADS (as illustrated in Figure~\ref{fig:fig_ads_generic}) interprets the environment using real-time data from multiple sensors (e.g., cameras, RADAR, and LiDAR), predicts trajectories of traffic participants, plans routes and makes decisions using a combination of ML models and code logic~\cite{pengzi}.
Thus, multi-module ADS is much more preferred than driving models in practice since driving models cannot handle complex driving scenarios and may easily overfit~\cite{Lou2022-vg}.  
%Figure~\ref{fig:fig_ads_generic} shows a common architecture of multi-module ADS. 

The perception module is one of the most critical modules in ADS. The perception module processes and interprets data from multiple sensors using deep neural network (DNN) models, such as the YOLO obstacle detection model~\cite{yolo} for detecting obstacles in camera images. If a perception module produces poor detection results (i.e., failing to detect obstacles), the safety and performance of an ADS will be significantly impacted. 
However, only a handful of prior research work~\cite{avfuzz, adfuzz, multisensor_merge_error, ebadi2021} provides automated solutions to perform system-level testing of L4 multi-sensor multi-module ADS and none of them focuses on exposing weakness of perception module for further testing an ADS. 

%Hence, 

%\todo{need an update here}In comparison, much more efforts are on end-to-end (E2E) driving models (e.g., Udacity's Chauffeur~\cite{chauffeur} in \cite{deeptest, deeproad}) or single-functionality autonomous driving assistance systems (L0/L1 ADAS) such as Adaptive Cruise Control system). %Previous work varies on the types of the applied ADS~\cite{Lou2022-vg}: end-to-end driving models (e.g., udacity's  Chauffeur~\cite{chauffeur} in \cite{deeptest, deeproad} and Autumn in \cite{deeproad}) or multi-module driving systems (e.g., Apollo~\cite{apollo} and Autoware~\cite{autoware}). 

In this work, we propose and implement a fuzzing technique (namely \tool) that generates diverse driving scenarios\footnote{A driving scenario contains a map (i.e., road structure and traffic lights), and information (e.g., location, destination, speed, etc.) of the ego car and all traffic participants.} which are used to test a multi-module ADS. Such system-level testing with a focus on the perception module allows us to {\it identify weakness in perception and further assess the impacts of perception failure on the entire ADS}. 
With a list of seeded driving scenarios ({\tt DS} for short), \toolS applies specialized mutation operators to the selected seed {\tt DS} for generating new test {\tt DS}. \toolS bypasses the challenge of generating multi-modal sensor input by leveraging a high-fidelity simulator, following prior work~\cite{avfuzz,adfuzz,multisensor_merge_error, ebadi2021}. A high-fidelity simulator such as LGSVL~\cite{lgsvl-paper} or CARLA~\cite{carla} generates diverse sensor output given one specified driving scenario.
Furthermore, \toolS positions the generation of {\tt DS} as a search problem and guides the generation using run-time information from the perception module of an ADS as feedback. Leveraging diverse run-time information allows \toolS to tailor the generation process and produce test cases that satisfy various criteria. 
One example of such criteria could be generating {\tt DS} with more undetected obstacles, i.e., there exist obstacles in the generated {\tt DS} but an ADS fails to detect them. 

We implement a prototype of \toolS and integrated \toolS with LGSVL~\cite{lgsvl}. LGSVL is commonly used by prior testing work on ADS, such as \cite{avfuzz, adfuzz}. We applied \toolS on an industry-grade L4 multi-sensor multi-module ADS, i.e., Baidu's Apollo. Apollo Robotaxi~\cite{apollo_robotaxi} has been deployed in major cities of China. We experimented with two metrics to guide the test generation of \tool. Our experiments show that \toolS identifies weakness in Apollo's perception module. Furthermore, guided by different metrics, \toolS generates diverse sets of driving scenarios, i.e., test cases for ADS. Last, we investigated top-ranked test cases  (i.e., ranked by the level of danger) generated \tool and identified several severe issues of Apollo 7.0.

In short, this paper makes the following contributions:
\begin{itemize}
    \item We propose a novel idea that (1) leverages a high-fidelity simulator to find weakness in ADS perception, and (2) exploits the weakness in perception to test an ADS.
    \item We implement a prototype (\tool) to test a commercial grade Level 4 ADS (Apollo by Baidu) against an open-source autonomous vehicle simulation platform (LGSVL). %\tool is highly extensible, i.e., can easily be extended to work with other ADS.  
    \item We applied \toolS and found serious issues (e.g., collisions) in a Level 4 commercial grade ADS. 
    %\item We further extend the evaluation of \toolS to an end-to-end fashion, i.e., to what extent, the generated scenes of poor perception will lead to catastrophic consequences, such as collisions.
    % \item We release the data from this work. \footnote{Artifact from this work is released at https://github.com/myproxemail2022/sim-test-ads}
\end{itemize}

% Evaluating the effectiveness of test suites on DNN models is explored by Ma et al. \cite{c6} using the concept of mutation testing. However, no prior work has investigated how to re-evaluate and re-use test data for model evolution. For example, new models are generated after retraining for better performance. The test data generated on the old versions can be used again to evaluate the new models. However, test data may require updates and selection since the test data may not yield the best performance (i.e., coverage and effectiveness) in the updated AI models.

\section{Background}

\pa{Autonomous Driving Systems and Apollo.}
% \todo{add a paragraph about the definition of ADS and compositions}
Existing Autonomous Driving Systems have different levels, ranging from Level 1 (L1), which is commonly referred to as {\it autonomous driver assistance systems (ADAS)} where the system provides some functionalities to support the driver such as adaptive cruise control (ACC) or automatic emergency braking (AEB) to Level 4 (L4) where the car is mostly autonomous (e.g., while having GPS and map data) and the driver can regain control of the car if needed~\cite{SAE_J3016}. In particular, Apollo~\cite{apollo}, an open-source ADS developed by Baidu, is aiming to be a L4 ADS.

%(at level 4) allows fully autonomous driving without needing any intervention from the driver. On a vehicle operated by one of these systems, the drivers are not required to take over driving since it is possible that these vehicles are not equipped with pedals or steering wheels. While ADS will only operate the vehicle in when all requirements are met and will not operate otherwise, it is enough to support use cases such as local driverless taxi~\cite{SAE_J3016}.

Figure~\ref{fig:fig_ads_generic} shows a simplified and generic structure of ADS. In a nutshell, an ADS is often composed of four major modules: perception, prediction, planning, and control. 
The perception module is responsible for seeing and interpreting the environments and obstacles based on sensor data. When an ADS is running on the road, its sensors (e.g., LiDAR and camera) capture real-time data streams on the surrounding environment. The time-series data is fed into the perception module, which then tries to detect things, such as obstacles, lanes, and traffic lights, and their location. 

The output from the perception module (i.e., types of obstacles and their location) is then fed into the prediction module. The prediction module predicts the trajectory of the obstacles (e.g., where a pedestrian is heading). The trajectory data is then fed into the planning module, which plans how the ADS should move towards the destination (e.g., move around the pedestrian to avoid a collision). Finally, the control module sends signals to control the vehicle (e.g., steering and braking). 

In this paper, we study Apollo, a production-grade open-source ADS developed by Baidu~\cite{apollo}.
Since its release in April 2017, Baidu has released 11 versions of Apollo. At the time of writing, the most recent version is Apollo 8.0. 
The current configuration of Apollo that we are using consists of one LiDAR sensor and two camera sensors.
%Apollo, among others has run tests of their ADS on real roads such as in Shenzhen, China\cite{apollo_progress_reuters}.
Apollo has been deployed to provide autonomous taxi services with no human intervention in multiple cities~\cite{apollo-go}.

\begin{figure}[ht]
\centerline{\includegraphics[width=\columnwidth,keepaspectratio]{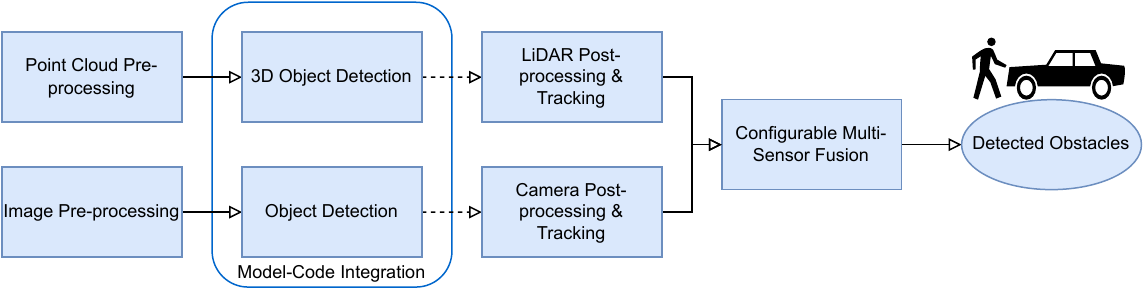}}
\caption{Obstacle Detection in the perception module of Apollo 7.0}
\label{fig:fig_apollo_perception}
\end{figure}
\pa{Obstacle Detection in the Perception Module of Apollo.} The perception module in Apollo is responsible for understanding the vehicle's surroundings which includes traffic lights, lanes, and obstacles~\cite{pengzi}.
In this paper, our goal is to test ADS by first finding problems in obstacle detection. 
Figure~\ref{fig:fig_apollo_perception} shows an overview of the obstacle detection process in Apollo 7.0, the newest version at the time of our experiment. 
% The perception module first receives data from multiple sensors. The data is then processed by code logic and ML models to understand the vehicle's surroundings such as the locations of traffic lights, lanes, and obstacles on and off the road.
% For example, 
% Apollo would need to filter the map data to know whether traffic lights are expected at an intersection. 
% Then, the perception module crops the camera input photo to only send the traffic light portion to ML models and code logic to identify the color of the light.
% Similarly, 
For obstacle detection, the raw data (point clouds, photos, etc.) from multiple types of sensors (LiDAR, camera, etc.) would be separately processed before being provided to the DNNs to detect and track these obstacles if they are within range.
There are two obstacle detection DNN models that are available for our experiment, one for LiDAR inputs and one for camera inputs.
For LiDAR obstacle detection, after the 3D point cloud pre-processing steps, the FCNN model~\cite{FCNN} receives a 2D grid, the cells of which represent characteristics of the original point cloud's ROI. The model then predicts the cell-wise obstacle attributes and these attributes are then used to segment out the obstacles.
For camera obstacle detection, the YOLO model~\cite{yolo} is responsible for 2D obstacle detection. It takes images as inputs and outputs the bounding box, type of the obstacle, etc.
After post-processing and tracking, the separate streams are fused using a predefined fusion strategy to obtain a final list of obstacles.

This list of obstacles shows the details of the obstacles detected by the perception module after fusion. 
For each obstacle, its output includes the id, 3D coordinates of its center, 3D coordinates of the points that form the bounding box, heading, velocity, acceleration, type, etc.
This combination of multiple sensor types allows Apollo to minimize discrepancies, allowing the sensors to complement each other (e.g., LiDAR does not have color information, yet is good at distance and velocity measurements; the camera have issues with depth perception but can detect colors).
However, challenges still remain as it is possible that only a portion of the obstacle is detected due to its size, orientation, and distance to the ego vehicle.
Ego vehicle is the name for the car that the ADS (in this case Apollo) is controlling.

%Thus, the detection results from these two sensor types are commonly fused together in ADS. For our purpose in this paper, we focus on the Perception Module's ability to detect obstacles. These related components are summarized in Fig. \ref{fig:fig_apollo_perception}

\pa{High-Fidelity Simulator for ADS.} 
In addition to physical road tests, simulation is a common approach adopted to test ADS in a real-world setting~\cite{surveypaper}.
Simulators~\cite{lgsvl, carla} create actual driving scenes and generate sensor data, which can then be used to train or test ADS. Simulators are significantly cheaper compared to physical road tests while still proving realistic testing environments. 
Johnson-Roberson et al.~\cite{10.1109/ICRA.2017.7989092} found that using simulators to train ADS may sometimes outperform training the ADS by actually running the vehicle on the road. Stocco et al.~\cite{stocco_emse23} show that simulator-generated data can be used as an alternative to real-world data.

In this paper, we use the LGSVL Simulator~\cite{lgsvl, lgsvl-paper}, which is a high-fidelity, well-known simulator built on Unity. LGSVL supports Apollo~\cite{apollo} for testing purposes. 
The simulator provides an out-of-the-box solution and a flexible Python API which allows for customization and expansion for multiple testing strategies. 
To generate a scene and the corresponding sensor data, the simulator generates the obstacles and a driving map based on a pre-defined configuration. Then, the simulated sensors would capture the scene and provide sensor data to appropriate ADS channels. Since the obstacles are generated in the simulator based on a given configuration, we can know the ground truth of the environment, and whether Apollo can detect all of the generated obstacles.
\section{Related Work}
%\todo{find the published citations instead of archive for as many as possible}
There are plenty of previous works on improving the safety of ADS through quality assurance. 
We organize related works based on the types of ADS being applied, from L1 ADAS to L4 ADS.
Due to the simple architecture of L1 ADAS, they can be represented as a single model or a system with separate modules for the single-sensor perception and feature such as ACC or AEB.
L2/L4 ADS can be further categorized based on system complexity, i.e., {\it multi-module} and {\it driving models}. 
% Before the application of DNN, ADS used to have limited functionalities, i.e., low level of automation, namely {\it non-DNN autonomous driving assistance systems (ADAS)}. 
Apollo (the test subject of this work) and Autoware~\cite{autoware} are the only two open-source L4 ADS, which are also multi-module ADS. Driving models only contain a single DNN model and often do not achieve a high level of automation, e.g., Udacity's Chauffeur~\cite{chauffeur} is L1.

%\pa{Scenario Generation for Testing ADS.}
%Althoff and Lutz~\cite{AlthoffLutz2018} and Klistchat and Alhoff~\cite{KlischatAlthoff2019} test the planning module by searching for scenarios where the driving area is limited.\jinqiu{rather theoretical work, it is possible to include this into testing planning}
%Laurent et al.~\cite{Laurent20201} proposes weight coverage representing aspects of driving and generate diverse scenarios for testing of a DA path planning system.

\pa{Testing L2/L4 Multi-Module Autonomous Driving Systems.} The closest related work to ours is a handful of previous work~\cite{avfuzz, adfuzz, multisensor_merge_error, ebadi2021, lawbreaker} that perform system-level fuzzing to an industry-grade multi-module ADS (i.e., Baidu's Apollo or comma.ai's openpilot~\cite{openpilot}).
 There are three key differences. (1)  {\it Different test module and focus.} Our focus is to find diverse driving scenarios that expose the weakness of the perception module and to study how such weakness further exposes safety concerns in the entire ADS. Li et al.~\cite{avfuzz} focus on safety violations in certain types of driving scenarios such as lane change. Zhong et al. \cite{adfuzz} aims to find traffic violations. Zhong et al.~\cite{multisensor_merge_error} detect issues in fusion logic (i.e., part of perception module). Ebadi et al.~\cite{ebadi2021} target the pedestrian detection and emergency braking system of Apollo. Sun et al.~\cite{lawbreaker} detect traffic law violations in the scenarios.
2) {\it Different fuzzing (mutations) and fitness function.} As we all have different test focuses, we employ different mutations and fitness functions for fuzzing. For example, Li et al.~\cite{avfuzz} perturb the driving behavior of traffic participants and search for such perturbations that minimize safety distance. %such that the ADS runs into violation. They use genetic algorithm to search for the perturbations to minimizes the safety potential of the ADS over its projected trajectory such that it runs into traffic violations.
Zhong et al.~\cite{adfuzz} mutate driving scenarios to find the ones causing traffic violations. 
Zhong et al.~\cite{multisensor_merge_error} also mutate driving scenarios and search for mutations that maximize fusion disagreements among sensors and maximize the correctness of one sensor. Sun et al.~\cite{lawbreaker} maximize specification coverage to find traffic law violations. %propose a simulation-based grey-box fuzzing framework to test the multi-sensor fusion logic of an advanced driver assistant system, and demonstrated that they can find caused by the multi-sensor fusion logic.
Differently, as we focus on the multi-sensor perception module, our focus is the mutation of obstacles (pedestrians, vehicles, their locations, and trajectories). We experiment with various fitness functions to guide the search process, i.e., coverage metrics of the DNN models, and obstacle detection results.
(3) {\it Black-box vs. grey-box.} We leverage run-time information (e.g., coverage metrics of a DNN model) from the test subject as feedback to guide the fuzzing, which makes \toolS closer to grey-box fuzzing. \cite{multisensor_merge_error} is also grey-box fuzzing as it uses internal data from the ADS under test while the others~\cite{avfuzz,adfuzz,ebadi2021} are black-box.
%However, there are a few key differences. 
%First, all three uses evolutionary-based fuzzing algorithms to find traffic violations using an objective function that minimizes the distance between the ego car and the leading/nearby obstacles to induce risky situations.
%Second, they have different targets for testings. In~\cite{avfuzz} since their perturbations are related to the parameters that decide the obstacles trajectories, the crashes are focused on the failure to predicts other obstacles trajectories, causing crashes in lane change scenarios. In~\cite{adfuzz}, they focuses on known, specific crashes defined in a previous work. In~\cite{multisensor_merge_error}, they focuses on multi-fusion merge errors of the perception module.

%Ebadi et al.~\cite{ebadi2021} uses Genetics Algorithm to search for test cases that generate a scenario where the ADS fails to stop for pedestrian.\jinqiu{this should be with the three:av-fuzzer, traffic violation, and fusion bug}

%Simulation-based testing is also applied for module level testing of ADS to by-pass the perception module which allow the following work to focus on the planning module.
In addition to system-level fuzzing, there are also research efforts on testing {\em a single module} in an ADS. %The closest one is 
Arcaini et al.~\cite{Arcaini2021} search for scenarios where specific driving patterns occur over a long duration to assess the correctness of the planning module.
Han and Zhou~\cite{metamorphic_prediction} leverage metamorphic testing to detect unavoidable collisions due to bugs in Apollo's planning and prediction modules. %generate obstacles to create collisions with the ego vehicle, finding unavoidable collisions.
Calo et al.~\cite{Calo20201} define and search for collisions that can be avoided due to issues in the planning module of a commercial ADS.
The closest work is by Zhou et al. \cite{metamorphic_lidar} where they test one type of sensor (i.e., LiDAR) in %the LiDAR-perception of 
Apollo by applying metamorphic testing. % on real-world LiDAR dataset. 
To the best of our knowledge, we are the first work that aims to test a multi-sensor perception (like the one in Apollo which includes both camera and LiDAR). % from an end-to-end system perspective. 
%\todo{this sentence's logic disconnects from the next}Part of our tool can be viewed as to apply fuzz testing to the multi-sensor perception of Apollo. 
Moreover, \toolS does not test the perception alone but also exercises the entire Apollo to assess whether the weakness of perception continues to impact the remaining modules of Apollo. 

\pa{Testing Driving Models.} A driving model (e.g., Chauffeur~\cite{chauffeur}) takes camera images and generates control decisions (e.g., steer angle) directly using one DNN model. Compared to multi-module ADS, a driving model is much more simplified, and testing a driving model is comparable to testing a single DNN model. So prior work that tests DNN models such as~\cite{deepxplore} can be applied to test such driving models.
%For LiDAR obstacles detection in the perception module, Zhou and Sun~\cite{metamorphic_lidar} test LiDAR obstacles detection for changes in obstacles detected by adding noises to the LiDAR input outside of the region-of-interest.
%Kutila et al.~\cite{Kutila2016, Kutila2018} find that the range of LiDAR sensors in ADS degrades during harsh weather and sensors with longer wavelengths perform better under such conditions.
Specialized techniques~\cite{deeptest, deeproad, asFault, selfOracle, combinatorial} are designed to test driving models. 
%Pei et al.~\cite{deepxplore} design neuron coverage-guided fuzz testing to test DNNs by activating rarely activated neurons.
Tian et al.~\cite{deeptest} apply weather transformations to real-world data sets and utilize neuron coverage of the underlying DNN model to guide the input generation. %metamorphic-based transformations representing various weather conditions to test ADS. 
Zhang et al.~\cite{deeproad} utilize GAN to transform images under extreme weather conditions for testing driving models. %normal driving condition to those with various weather conditions for metamorphic testing.
%However, given how each sensor-model combination has a different data input type, e.g., photos, point clouds, etc. tests of obstacle detection based on one type of sensor is not transferable to that of another sensor type.
%Moreover, given how most level 4 ADS eventually fuse the results of obstacle detection from multiple sensor sources, it is difficult to verify whether the bugs detected from these testing approaches are filtered out after the fusion process.
Such specialized techniques designed for driving models cannot be applied to test multi-module ADS directly. Multi-module ADS employs multiple sensors and applying consistent transformations on multi-modality sensor input remains challenging.
\toolS leverages a high-fidelity simulator to overcome the challenge of generating multi-modality sensor data. It remains future work to incorporate the advanced weather transformations in a high-fidelity simulator.
%Our work is different from the following aspects. First, \toolS is a general testing tool and can be applied to test various types of sensors, LiDAR, camera, and camera-LiDAR fusion. Second, \toolS overcomes the oracle challenge by leveraging high-fidelity simulators. Third, \toolS performs system-level testing an ADS, by utilizing weakness in perception, and thus is different from previous work focusing on different types of ADS defects. 

\pa{Testing L1 Autonomous Driving Assistance Systems.} Researchers propose various techniques to test the safety of L1 ADAS~\cite{Dreossi2017, Abdessalem2016, Abdessalem2018, Abdessalem2020, Belerovic, Koren2018, Akagi2019, Corso2019}. 
However, such techniques proposed for L1 ADAS cannot be applied to L4 multi-sensor multi-module ADS (such as Baidu's Apollo) due to the complex architecture of L4 ADS, i.e., multiple types of sensors for accurate perception and modules for comprehensive functionalities. This architecture provides safety checks and handles single faulty sensor scenarios which renders the techniques that target L1 ADAS obsolete for L4 ADS. For example, the bugs found via testing techniques that focus on an ADAS that decides whether to stop based on single camera images are very unlikely to work on an L4 ADS because in the L4 ADS, there would be at least the image taken by another camera or the point cloud taken by LiDAR that does not encounter the same problem. 
The works~\cite{Zhao_2017, Huang2016, Huang2017} that propose techniques to extract driving scenarios from video recordings, could benefit our work, i.e., providing a more diverse and real-world initial population of driving scenarios.

\pa{Adversarial attacks on Autonomous Driving Systems.} Many adversarial attack techniques have been proposed in recent years. We describe the attacks on the perception module of an ADS below. We refer to a recent survey on a comprehensive list of papers about attacks on ADS~\cite{attack_survey}. Most works propose adversarial attacks that target a single-sensor (LiDAR or camera) perception module~\cite{physical_stickers_attack_camera_disappear_stop_sign, Eykholt2018, ShapeShifter, seeing_isnt_believing, Illusion_and_Dazzle, Rubaiyat,lidar-attack}.
Cao et al.~\cite{invisibleobject} propose the first technique that can generate adversarial 3D obstacles that can fool both LiDAR and camera obstacle detection.
Different from previous adversarial attacks, we focus on testing a multi-module ADS and propose an automated testing framework without assuming there is a malicious attacker. % standby to fool the perception module.

\begin{figure*}
%\centerline{\includegraphics[width=\textwidth,height=\textheight,keepaspectratio]{figures/How_Simulation_Fuzz_workswith_Apollo.pdf}}
\centerline{\includegraphics[width=\textwidth,height=\textheight,keepaspectratio]{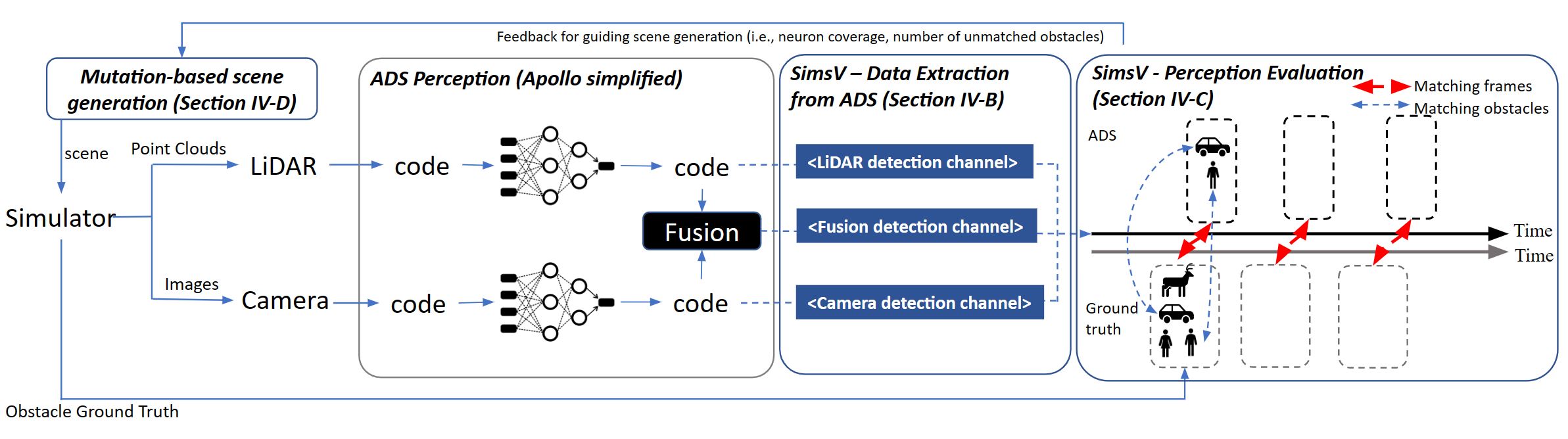}}
\caption{An overview of \tool}
\label{fig:overview}
\end{figure*}
\section{\tool: Methodology and Implementation}
In this section, we describe the methodology and implementation of \toolS in detail. For each component of \tool, we first describe its methodology, which is independent of ADS and simulator. Then we describe our implementation of \toolS for Apollo (an industry-grade ADS) and LGSVL (a high-fidelity simulator). 
%Our prototype of \toolS is independent of the ADS and self-driving simulation platform, and can be extended to support other ADS and driving simulation platforms.

%our automated framework for measurements of the systems and detect errors. 
\subsection{An Overview of \tool}
Figure \ref{fig:overview} illustrates an overview of \tool and its interactions with ADS and the simulator. \toolS contains three main components, namely data extraction from ADS,  scene generation, and perception evaluation. 
Given a set of initial seeds, \toolS adapts an evolutionary fuzzing algorithm for generating diverse driving scenes. The generated driving scenes are simulated by a simulator and are further exercised by the ADS under test. 
%The selected simulator will then be exercised to generate driving scenes given the generated parameters. 
Such driving scenes are of high fidelity, including LiDAR data, images for the camera, all the obstacles in the simulated environment, etc. 
%The ADS in the simulator perceives the simulated environment, , plans its trajectory and makes critical driving decisions. 
\toolS monitors the relevant communication channels of the ADS under test and extracts run-time information from the perception module for later use: activated neurons in the DNN models for neuron coverage calculation, and DNN model outputs (Section~\ref{sec:perception}). 

\subsection{Data Extraction from ADS}
\label{sec:data_extraction}
\pa{Methodology.} \toolS utilizes run-time information of the ADS under test to guide the fuzzing process, similar to grey-box fuzzing in literature~\cite{fuzzing}. We experiment with two metrics in this paper for guiding the fuzzing. This component of \toolS extracts two types of real-time information from an ADS: (1) obstacle detection results from solo-camera, solo-LiDAR, and LiDAR-camera fusion; and (2) coverage information of the CNN model for LiDAR obstacle detection and the YOLO model for camera obstacle detection). 
 
 \pa{Implementation.}
 To obtain (1) obstacle detection results, \toolS 
 is implemented to work with Apollo's underlying real-time framework, CyberRT~\cite{cyberrt} to monitor communication channels representing Apollo's obstacle detection output. Typically, the perception module receives 10 frames/s per sensor. For LiDAR-camera fusion, we monitor \texttt{/apollo/perception/obstacles}.
 For LiDAR-only obstacle detection, we expose this data in a new channel called \texttt{/apollo/perception/obstacles\_lidar},
 and for camera-only obstacle detection, we re-enable the \texttt{$\slash$perception$\slash$obstacles} channel. 
 Similar to other real-time systems, ADS relies on communication channels to exchange internal messages. For example, when the perception module publishes obstacles detected messages, the prediction module is notified of those obstacles so it can calculate their trajectories. Such a publish-subscribe architecture allows \tool to monitor Apollo's internal communication in a {\it lightweight} and {\it non-intrusive} manner.
 
% Autonomous driving systems, which is one type of real-time systems, commonly adapt an architecture of channel publish/subscribing for internal message communication. 
% For testing Apollo, we implemented \toolS to work with its underlying real-time framework, CyberRT~\cite{cyberrt}. 
% Note that, \toolS may be adopted to analyze other real-time frameworks when testing other ADS (e.g., the \textit{ROS 2} framework that is used by \textit{Autoware}). 
% Typically, the perception module receives 10 frames/s per sensor for processing.  
%This part is conceptually similar for other real-time frameworks adopted by other ADS, such as \textit{ROS 2} by \textit{Autoware}. 

 To obtain (2) coverage information of DL models, \toolS instruments the two Apollo-customized Caffe DNN models in Apollo and records the activated neurons per simulation: one for LiDAR and the other one for the camera. Following previous work~\cite{deepxplore}, \toolS considers a neuron is activated when the mean of the output is greater than a threshold. For each data input frame received, we measure the neuron activation from all the intermediate layers of the model processing the input. The neuron activation of each round is the combined activation of all the input frames generated by the simulator in that round. 
 Since the computation of the neuron coverage is time-consuming compared to Apollo's normal pipeline, if Apollo is run at real-time speed, it would drop input frames causing incorrect detection results. 
 To accommodate this issue, we slow down the time of the simulator (so that it takes five seconds of clock time for one second in the simulator to unfold) when we need to measure the neuron coverage.
%This component consist of various functionality that is embedded in Apollo to extract and measure different types of information. It records the neuron outputs and monitors the resulting obstacle detection data from different sensors and the resulting fusion. It will match these detect obstacles to the ground truth and relay all these information to the Fuzzing Component.

%The functionality of this component in relation to the framework is shown in shown in Fig. \ref{fig_measurement_component}
\begin{comment}
\subsubsection{Obstacle Detection Data Extraction \& Matching to Ground Truth}
\begin{figure}[htbp]
\centerline{\includegraphics[width=\columnwidth,keepaspectratio]{figures/Detection_Rate.pdf}}
\caption{Data Flow of Measurement Component}
\label{fig_measurement_component}
\end{figure}
\pa{Monitoring perception results in Apollo.}
In this step, the component would monitor the data channels in Apollo. These channels are streams of data containing the obstacles detected from sensors input (10-15 frames/second). 
These individual channels which contains the obstacle detection data are monitored

\jinqiu{need a figure here: from which channels channels }\triet{This is already in Fig. 2. Should we add another?}
\begin{itemize}
    \item The obstacles ground truth in range channel (from Simulator Component)
    \item The fusion obstacles perception output channel (from Apollo)
    % \item The LiDAR obstacle detection channel (from Apollo)
    \item The camera obstacle detection channel (from Apollo)
\end{itemize}
\end{comment}

\subsection{Perception Evaluation}
\label{sec:perception}
The perception evaluation module in \tool receives and compares: (1) Apollo's obstacle detection results from one of the three communication channels (Section~\ref{sec:data_extraction}); and (2) ground truth from the simulator.
For both sides, the data frames (e.g., messages in the channels monitored by \tool) are received continuously. For example, a simulation run of five seconds will result in a total of 50 data frames with an average of 10 frames/s.
\begin{figure}[ht]
    \centering
    \includegraphics[width=\columnwidth]{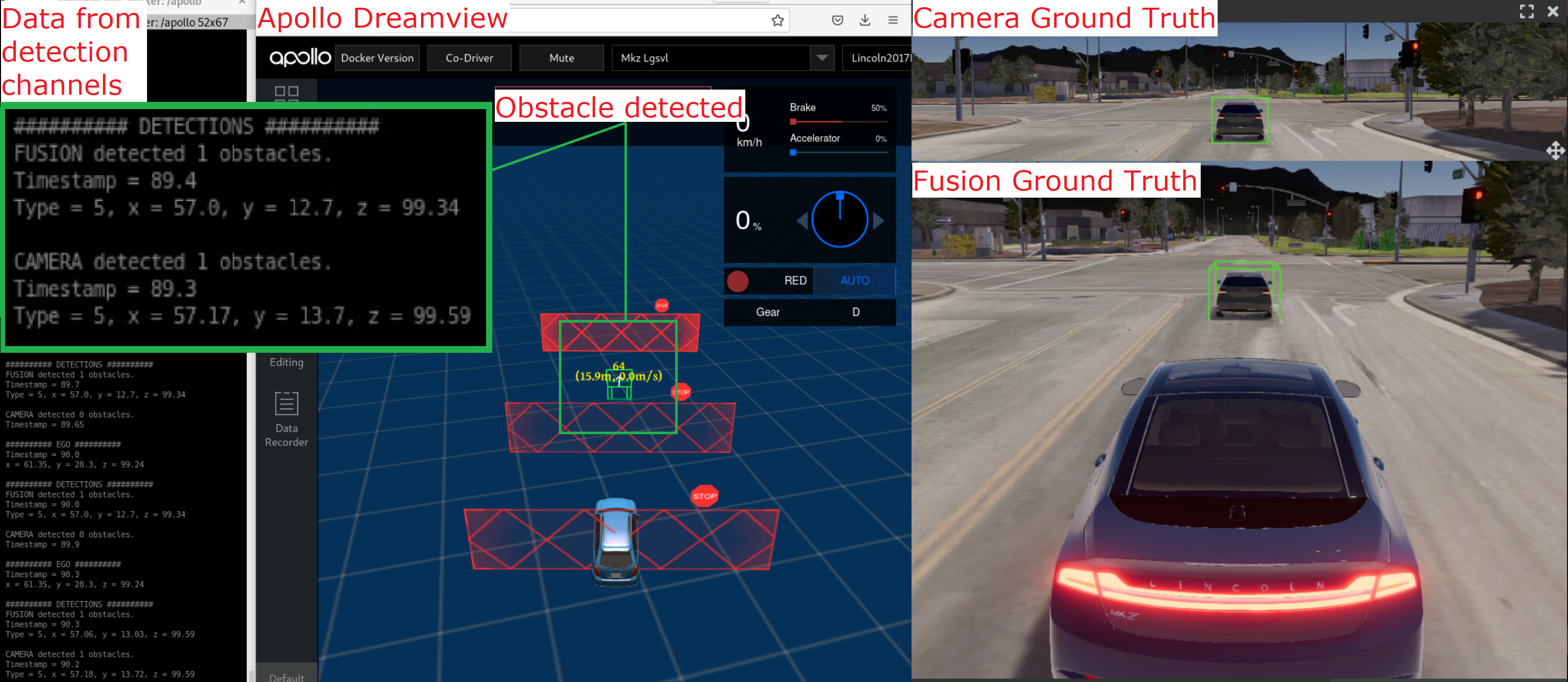}
    \caption{\toolS compares the obstacle detection results from Apollo with the ground truth. Enlarged and highlighted in green is an abstracted message received from detection, which includes the detected obstacles and their attributes (i.e., the position of the center, velocity, etc.). Three windows are displayed in the figure. From left to right are windows showing (1) obstacle detection messages, (2) obstacles detected visualized in Dreamview, and (3) obstacles as they appear in LGSVL.
    % \todo{@Triet, edit this figure in powerpoint, the detection result is unreadable}
    }
    \label{fig:detection_frame}
\end{figure}
Each message of obstacle detection contains the time stamp of when obstacles were detected and a list of detected obstacles, each containing its position, velocity, etc.
Figure~\ref{fig:detection_frame} shows an example when the vehicle in front of the ego car is detected. The detected obstacle details can be seen in the enlarged, highlighted in green message in the left window and the visualization in the Dreamview window in the center. This detection matches the ground truth as shown in the simulator window on the right. Dreamview~\cite{dreamview} is Apollo's built-in tool to visualize its modules output, e.g., the ego vehicle's localization, planning trajectory, obstacles detected, etc.

% \todo{describe the example message for obstacle detection: an obstacle (car) in front of the ego car.}

\toolS performs the comparison through a two-step matching (as illustrated in the right-most box in  Figure~\ref{fig:overview}).

% \pa{Step 1: Matching data frames.} \toolS aligns the two sets of data frames based on time stamps.
% Due to the sequential nature of the data frames (i.e., data frames containing the ground truth from the simulator have earlier time stamps than those representing obstacle detection from Apollo), 
% \toolS tries to align the matching frames whose time stamps are less than a predefined threshold (currently 30 ms).
\pa{Step 1: Matching data frames.} Using time stamps, \toolS matches the data frames containing the ground truth from the simulator with the data frames containing the detected obstacles via different sensors from Apollo. 
This can be achieved by utilizing two characteristics of the simulator and Apollo.
First, for each data frame containing the ground truth, there is only one data frame containing the obstacles detected for each sensor type. Second, the data frames containing the detected obstacles are generated after the data frames containing the ground truth obstacles, hence, they always have later time stamps and arrive sequentially. These two characteristics allow us to match a data frame containing the ground truth directly with its counterpart containing the detected obstacles.

\pa{Step 2: Matching obstacles.} For each pair of matching frames, we use the Hungarian matching algorithm to match obstacles that are: (1) detected by Apollo; and (2) from the simulator's ground truth. Obstacles are divided into subsets based on categories (e.g., vehicles and pedestrians) before they are matched to avoid crossovers (e.g., a pedestrian should not be matched with a car). For this N-to-M matching problem, the distance between each pair is defined as their Manhattan distance. The Hungarian matching algorithm is utilized to find the assignment that minimizes the sum of the distances of all the pairs in the assignment. 
%This allows the computation of the obstacle detection rate of each type of sensor and the fusion strategy.
%Obstacles of different types or with a x difference of greater than 4 meters and y difference of greater than 1 are not matched to one another.

Once the matching is complete, we define the following three metrics to measure the performance of a perception module on obstacle detection.
\begin{itemize}
\item {\bf Precision: } Per data frame, the percentage of detected obstacles by an ADS that are in the ground truth.
    For example, for the first two frames matched in the right-most box in Figure~\ref{fig:overview}, one pedestrian and one car are matched between ADS detection and ground truth (linked by blue dashed lines), the precision of this data frame is 100\% (2/2). Low precision means detection of non-existent obstacles, which may lead to unnecessary stops~\cite{tooafraidtodrive}.%\todo{adverse consequences of low precision}\triet{I think this cause unnecessary stop}
    
\item {\bf Recall: } Per data frame, the percentage of obstacles in the ground truth that are detected by an ADS. For example, the recall for the first data frame in Fig.~\ref{fig:overview} (the right-most box) is 50\% (2/4). Frames with low recall can have dire consequences for autonomous driving including collisions or failure to compute trajectories. The most straightforward is the ego car will collide with an obstacle in front of it if the obstacle cannot be detected.

\item {\bf Perception rate: } Per ground-truth obstacle over a period of time (e.g., one simulation round), the percentage of data frames that this obstacle is detected. For example, imagine the three data frames in Figure~\ref{fig:overview} (the rightmost box) occur in 0.3 seconds, the obstacle car is in two out of the three data frames, the perception rate of the obstacle car in the 0.3 seconds would be 66.7\% (2/3). A high perception rate means that an obstacle is reliably detected by an ADS, which is very important to satisfy the safety requirement of ADS.
\end{itemize}
%This data is sent to the Fuzzing Component for further analysis.

\subsection{Mutation-based scene generation}
Given a seed scene, \toolS supports six mutation operators that transform the present obstacles, e.g., adding or removing one obstacle. This fuzzing process can either be random or guided (i.e., transformations are performed to maximize certain metrics) manners. In this paper, we experimented with guiding using neuron coverage, which is commonly used in prior work~\cite{deepxplore, dlfuzz}.
%So far, we provide implementations of several\jinqiu{XXX} metrics for guiding the fuzzing process.
%\toolS overcomes the test oracle challenge by utilizing a driving simulator, i.e., ground $ADS_{perception}$(simulator(ground truth)) == ground truth  test oracle is pre-defined as the input to the simulator for simulating 

\begin{algorithm}[ht]
%\begin{algorithmic} 
\KwIn{Initial seeds {\it $I$}}
\KwIn{Fitness function {\it $f$}}
\KwOut{Generated driving scenarios {\it $genSC$}}
 Push all initial seeds in \textit{I} to Queue {\it Q}\;
 \textit{genSC} = $\emptyset$\;
 \While{Q is not empty}{
    \textit{scn} = \textit{Q}.dequeue()\;
    Randomly pick a mutation operator \textit{Op}\;
    Randomly pick parameters for \textit{Op} as \textit{paras}\;
    \textit{newScn} = Mutate({\it scn, Op, paras})\;
    \eIf{f\textit{(newScn, scn)} $>$ 0}{
    \textit{genSC} = \textit{genSC} $\cup$ \textit{newScn}\;
    Q.enqueue(newScn)\;
    }{
    Q.enqueue(scn)\;}
 } 
 
\caption{Mutation-based scene generation\label{alg:mutation}. Can be guided or non-guided.}
\end{algorithm}
% \begin{table}[ht]
%     \caption{\todo{fixme for consistency}Different mutation operators and their parameters.}

%     \centering
%     \begin{tabular}{L{1cm}L{4cm}L{2.2cm}}
%     \toprule
%      {\centering Mutation}  & Description & Parameters\\
%      {\centering Operator}  &             &           \\\hline
%       \multirow{2}{*}{Add}  & \multirow{2}{4cm}{Adds a random obstacle at a random position within the detection range of the ADS}                         & - an obstacle\\
%                             &             & - a position in the scene\\[0.2cm]\hline
%        Remove               & Removes one randomly-chosen obstacle from the scene            & - an obstacle from the scene\\[0.2cm]\hline
%        Swap                 & Swaps the locations of two random obstacles from the scene           & - two obstacles from the scene\\[0.2cm]\hline
%        \multirow{2}{*}{Move}& \multirow{2}{4cm}{Moves the location of a randomly chosen obstacle either towards north, south, east, or west (map direction) by one meter (map coordinates)}&  - an obstacle from the scene\\
%                            &              & - a new location\\[0.4cm]\hline
%        \multirow{2}{0.8cm}{Modify Velocity}& \multirow{2}{4cm}{Modifies the velocity of a randomly chosen obstacle from the scene} & - an obstacle from the scene\\
%        & & - a random velocity \\[0cm]\hline
%        Rotate &  Rotates the direction a randomly chosen obstacle is facing by $90^\circ$ clockwise & - an obstacle\\\bottomrule 
%     \end{tabular}
%     \label{tab:my_label}
% \end{table}

\begin{table}[ht]
    \caption{Different mutation operators and their parameters.}
    \centering
    \begin{tabular}{L{1.5cm}L{8cm}L{4.5cm}}
    \toprule
    Mutation Operator    & Description & Parameters\\\hline
    Add                  & Adds a random obstacle at a random position within the      & - a randomly generated obstacle\\
                         & detection range of the ego car                              & - a position in the scene\\[0.1cm]\hline
    Remove               & Removes one randomly-chosen obstacle from the scene         & - an obstacle from the scene\\[0.1cm]\hline
    Swap                 & Swaps the location of two randomly chosen obstacles from the& - two obstacles from the scene\\
                         & scene                                                       & \\[0.1cm]\hline
    Move                 & Moves a randomly chosen obstacle either north, south, east, & - an obstacle from the scene\\
                         & or west (map direction) by one meter (map coordinates)      & - the new location\\[0.1cm]\hline
    Modify               & Modifies the velocity of a randomly chosen obstacle from the& - an obstacle from the scene\\
    Velocity             & scene                                                       & - a random velocity (including 0) \\[0.1cm]\hline
    Rotate               &  Rotates the direction a randomly chosen obstacle is facing by $90^\circ$ clockwise & - an obstacle from the scene\\\bottomrule 
    \end{tabular}
    \label{tab:my_label}
\end{table}

\pa{Fuzzing process.}
\toolS adapts an evolutionary mutation-based fuzzing process to drive the generation of driving scenes towards certain goals, i.e., minimizing precision, and activating rarely-activated neurons. The fuzzing process is repeated in a loop up to a specified number of rounds.

Algorithm~\ref{alg:mutation} describes the fuzzing algorithm in detail. \toolS performs a breadth-first search (i.e., the while loop starting line 4) through a queue for maintaining a pool of candidate seeds. \toolS applies mutation operators (lines 6-8) to one seed scene and evaluates if the synthesized scene would outperform the seed scene measured by certain metrics, i.e., the use of fitness function in lines 9-12.
One such fitness function is the number of newly activated neurons. If the mutated scene activates previously inactive neurons, it is kept for future mutations. 
%This evaluation using fitness function requires 
Given the gigantic and sparse search space (i.e., tremendous combinations of obstacles being placed), evolutionary fuzzing allows a targeted search in the space and yields more interesting findings in a limited time. 

\pa{Mutation operators on driving scenes.} \toolS leverages a driving simulator for generating realistic sensor data. Hence \toolS proposes mutation operators that can be applied to driving scenes. Such mutation operators modify obstacles so that such scene transformations are substantial, non-trivial (compared with changing pixels or bits of images), and realistic since the mutation changes deal with the obstacles' position, velocity, and orientation.
In total \toolS uses six mutation operators (Table~\ref{tab:my_label}). Each mutation operator requires at least one parameter. Some parameters are randomly generated. For example, when applying the \texttt{add} mutation operator, an obstacle needs to be randomly generated before being added to the scene. We use the existing agents in the LGSVL simulator as obstacles. For example, \svl~provides 11 prototypes for pedestrians (e.g., different heights, clothes, etc.), one prototype for turkey, and one for deer. For vehicles, we use five models representing common vehicle classes on the road. 

\pa{Fitness function.} The design of a fitness function decides what type of test cases are generated through fuzzing. In this paper, we experiment with two fitness functions, both of which require run-time information from Apollo's perception. The first fitness function is based on comparing activated neurons between the newly generated scene (i.e., $AN_{muScn}$) and the original one: 
$\{n \mid n \in AN_{muScn} \land n \notin AN_{origScn}\}$. %Note that each round contains N frames depending on the time of this round, and the activated neurons of this round is the
Each simulation round contains multiple frames depending on the running time, so the set of activated neurons per round is the union of activated neurons of all the frames. 
$AN_{origScn} = AN_{f1} \cup AN_{f2} \dotsb \cup AN_{fn}$.
The second fitness function we experimented with is the average number of undetected obstacles by the ADS. For each frame, the number of undetected obstacles is recorded (Section~\ref{sec:perception}). Then an average is calculated for each simulation round driven by one driving scene.

\section{Evaluation}
This section first describes our evaluation plan of \tool. We describe our study subjects, set up, and methods to evaluate \tool. Then, we present our evaluation results. 

% \subsection*{Experiment Subjects and Setup} 

Below we detail our experiment subjects, including the ADS (Apollo), simulator (\svl), the configurations of \tool, and computing resources that we used in this experiment.

\pa{Apollo.} We performed the evaluation of \toolS on Apollo v7.0, which was released on December 28, 2021. Apollo v7.0 adapts multi-sensor fusion, using camera, LiDAR, and radar for perception. 
However, even the most compatible simulator (i.e., \svlS) with Apollo does not support radar for Apollo in a simulated way.
Hence, our evaluation focuses on {\it solo} LiDAR, {\it solo} camera, and camera-LiDAR fusion detection. 
%Our implementation of \toolS does support testing the multi-sensor fusion by monitoring the fusion detection channel. It remains as future work to apply \toolS on Apollo v5.5 for testing the multi-sensor fusion.
%\todo{talk about the SMOKE model? do we test that or not?}\triet{We have not test this}

\pa{\svl.} We utilize LGSVL 2.2 (an open-source driving simulator) due to its compatibility with Apollo. We implemented \tool's mutation operators by leveraging APIs of LGSVL. 
LGSVL takes each generated scene by \toolS as an initial scene to begin a simulated run with Apollo: All the obstacles will follow traffic rules and move based on their individually set velocity.

% \todo{describe scene minus obstacles, we only use one I believe}
% LGSVL is shipped with a default scene. 
For our experiments, we use the default scene that came with LGSVL, namely Borregas Avenue. This scene is a digital twin of a real-world street block of the same name in Sunnyvale, CA. It features two-lane roads as well as one traffic light intersection and one stop sign intersection~\cite{lgsvl}.
% \todo{screenshots and description of the scene}
In this evaluation, we set each of the simulated runs to last five seconds in RQ1 and 45 seconds or until the ego car arrives at a preset destination for RQ2. These numbers are selected because RQ1 is meant to cover a wide breadth of scenarios to identify issues and RQ2 is meant to perform in-depth analysis of unsafe scenarios identified in RQ1.

\pa{\toolS Configurations.} The initial seeds of \toolS in this experiment are ten distinct scenes: Each scene includes one randomly generated obstacle within the ego car detection range. 
%\todo{quality detection: 80 meters for car and pedestrian, claimed in the documentation, what is the link to the documentation, }
% \todo{neuron activation threshold}.\triet{added the following} 
For the neuron activation threshold, we use the threshold value of 0.1 to compare the neuron output for activation.

\pa{Computing Environment.}
% \todo{add lambda spec}\triet{added}
All experiments are run on a server with AMD Ryzen 16-core CPU, 256 GiB of memory, and two NVIDIA GeForce RTX 2090. The installed OS is Ubuntu 20.04.3 LTS.

%\section{Evaluation Results}
\subsection*{RQ1: \RQOne}
\label{eval:random_fuzz}
\pa{Motivation.}  
%We believe that \toolS can still reveal unique weakness in Apollo's perception 
Even though a commercial-grade ADS such as Apollo has been thoroughly tested, we wonder if simulation-based testing can still identify unique weaknesses in Apollo's perception. We are interested in trying a large number of possible scenarios, each for a very short amount of time to find interesting cases before later in-depth analysis.

\pa{Method.} We apply \toolS(neuron-guided) to test Apollo's perception module with only input from: solo camera, solo LiDAR, and camera-LiDAR fusion, each for 300 rounds. For camera-LiDAR fusion, we also utilize the average undetected obstacles within the detection range fitness function for the guidance of \toolS for 300 rounds.
% for a total of 900 rounds over three sources.

For solo camera and solo LiDAR, we set \toolS to monitor the channels for the individual detection results from the camera and LiDAR. When testing each sensor (camera or LiDAR), the corresponding DNN model is used for neuron coverage guidance.
For camera-LiDAR fusion, since there are two models involved for obstacle detection, as in the solo cases. We use the newly activated neurons from both models for the fitness function.
%In general, a perception module in an ADS enables multiple sensors, i.e., \lidar, camera and radar. The subject ADS of this work--Apollo utilizes \lidarS (i.e., three LiDARs with different resolutions) and camera, and fuses the detection results (as shown in Figure~\todo{XXX}). In this work, we applied \toolS to the various settings of Apollo Perception, namely 1) LiDAR 16; 2) LiDAR 64; 3) LiDAR 128; and 4) camera. Our subjects under test (SUTs) of Apollo are not limited to the DL models employed for obstacle detection from sensor data, but also the source code of Apollo Perception that is responsible for receiving, pre-processing the data from sensors, and post-processing of detected obstacles data from the DL models before forwarding it to the next component in Apollo.

For each sensor type, using each generated scene by \tool as an initial scene, for each simulation round, we let the ego car by Apollo continue self-driving for five seconds.
In total, we ran \toolS for a period of eight hours and obtained a total of 300 simulation rounds with distinct initial scenes that are generated by \tool.
\begin{figure}
     \centering
     \includegraphics[width=\columnwidth,keepaspectratio]{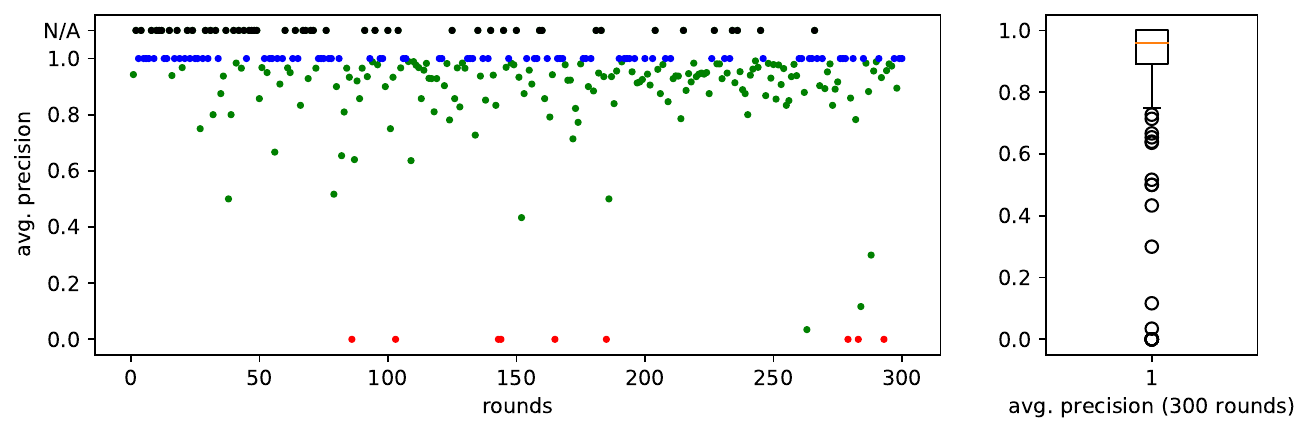}
    \includegraphics[width=\columnwidth,keepaspectratio]{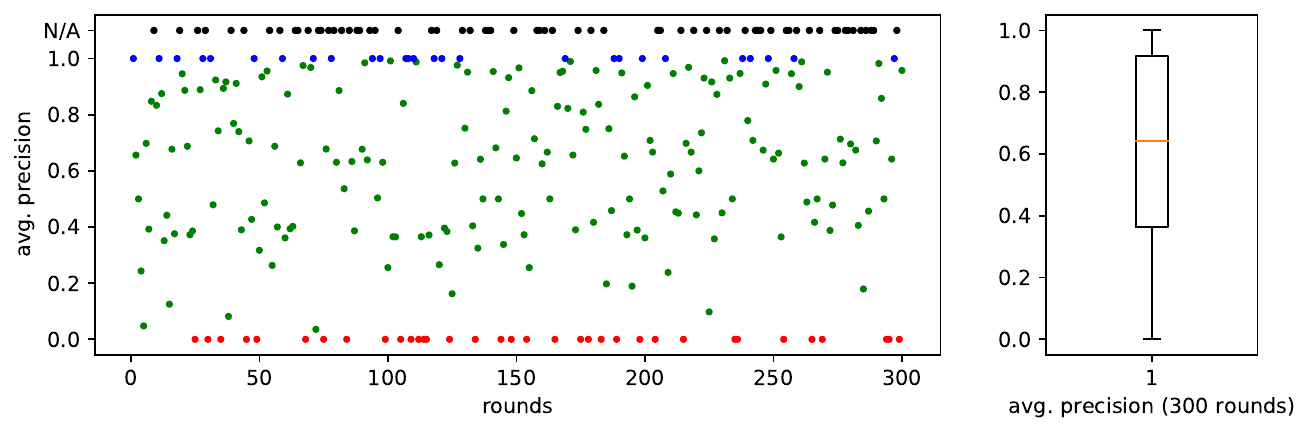}
    \includegraphics[width=\columnwidth, keepaspectratio]{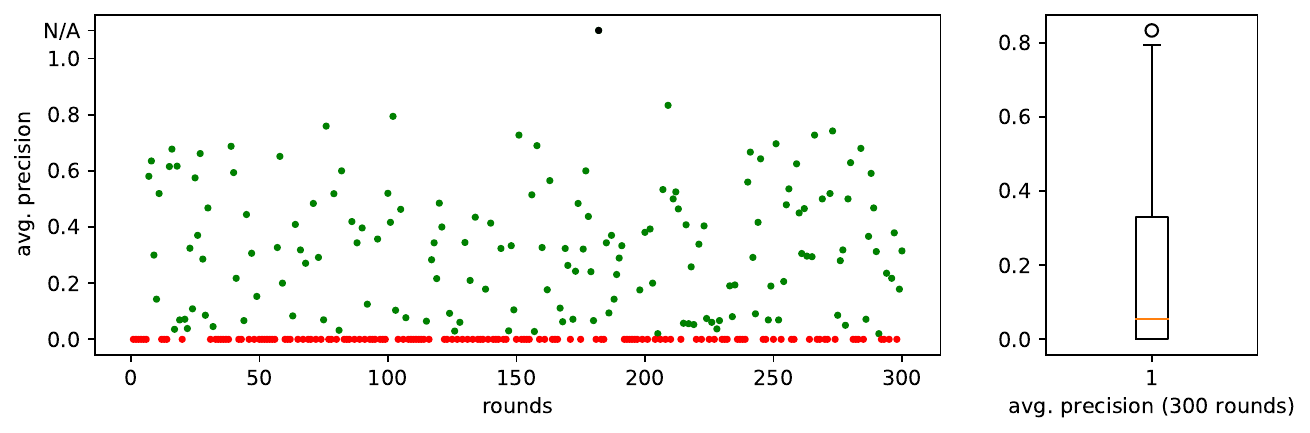}
     \caption{Average precision for 300 rounds by \toolS (neuron-coverage guided) for camera-LiDAR fusion (the top row), solo LiDAR (the middle row), and solo camera (the bottom row).} 
     \label{fig:avg_precision_RQ1}
 \end{figure}
Each simulation round results in 50 data frames from the perception module. For each data frame, \toolS calculates a precision and a recall (Section~\ref{sec:perception}).
The perception rate is calculated for each obstacle in one simulation round, which indicates whether one obstacle can be reliably detected. Also, we monitor the position of obstacles with respect to the ego car to ensure that they are within the sensors' detection range and the perception rate is meaningful.
% , i.e., 50 point clouds and 50 camera images

\pa{Results.}
Across all 900 rounds of neuron-guided fuzzing from all sensors, the maximum distance between the ego car and the obstacles is 56 meters, which is within the detection range. For the number of rounds that we performed, the number of obstacles in the scenarios was between one and three obstacles.

Figure~\ref{fig:avg_precision_RQ1} depicts the average precision across all the 300 runs for camera-LiDAR fusion and solo LiDAR (solo camera is in the artifact due to space concern). "N/A" stands for the rounds that Apollo perception detects 0 obstacles, then precision cannot be calculated.
Camera-LiDAR fusion has a much higher precision compared with solo LiDAR, and the camera is much worse than fusion and solo LiDAR. 
Even for the best-performing fusion, \toolS still finds that in 15/300 rounds, the average precision is lower than 50\%.

\begin{figure}
    \centering
     \includegraphics[width=\columnwidth,keepaspectratio]{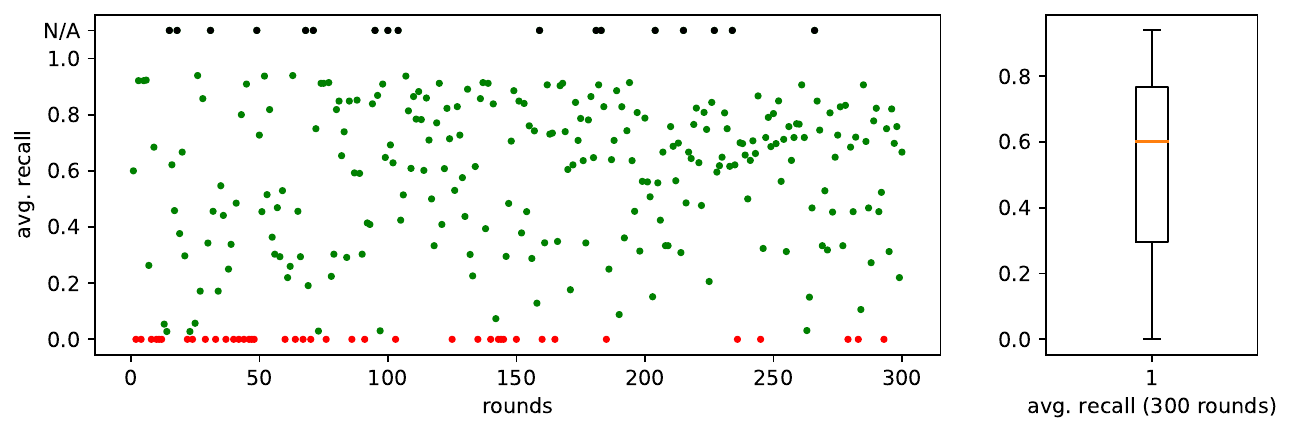}
    \includegraphics[width=\columnwidth,keepaspectratio]{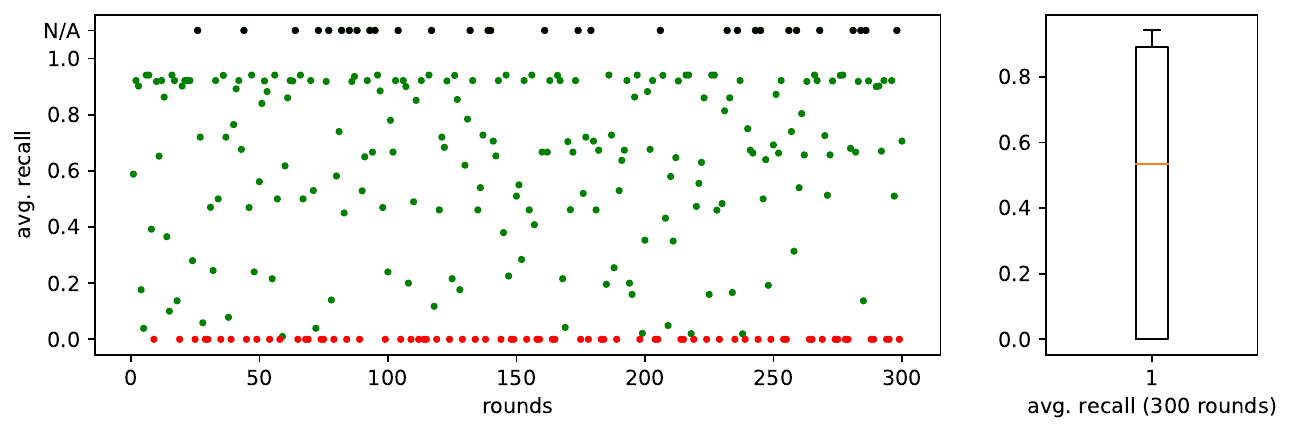}
    \includegraphics[width=\columnwidth,keepaspectratio]{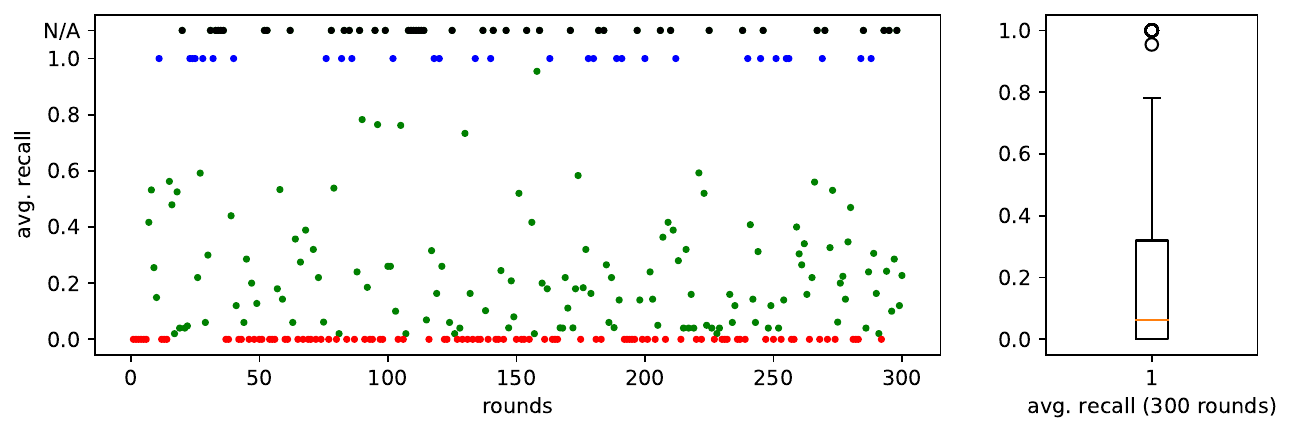}
     \caption{Average recall for 300 rounds by \toolS (neuron-coverage guided) for camera-LiDAR fusion (the top row), solo LiDAR (the middle row), and solo camera (the bottom row).\label{fig:recall_neuron}}
\end{figure}

Figure~\ref{fig:recall_neuron} shows the average recall across the 300 rounds. Camera-LiDAR fusion is still much better than solo LiDAR in terms of recall, however the improvement is limited compared to precision. The median of recall is 0.6 for camera-LiDAR, which means that for 50\% of the 300 rounds, at least six out of ten obstacles can not be detected by Apollo perception. Low recall is more problematic than low precision as low recall can significantly increase the chance of collision.    

\begin{figure}
    \centering
     \includegraphics[width=\columnwidth,keepaspectratio]{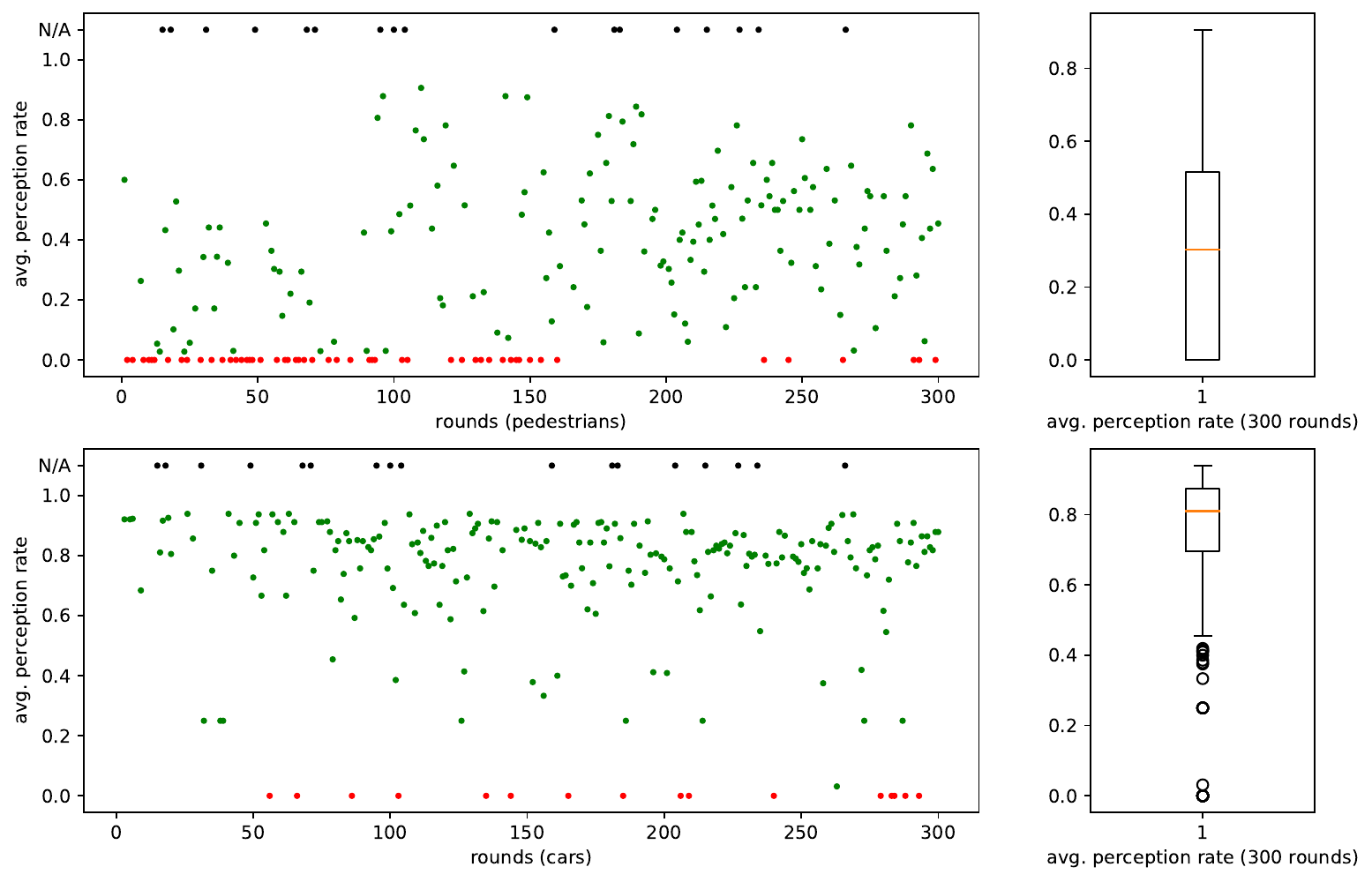}
     \caption{Average perception rate for 300 rounds by \toolS (neuron-coverage guided) for camera-LiDAR fusion. The top graph shows the average perception rate of pedestrians each round and the bottom graph shows the average perception rate of vehicles in the corresponding rounds.\label{fig:perception_rate_neuron}}
\end{figure}

Figure~\ref{fig:perception_rate_neuron} shows the average perception rate across the 300 rounds for fusion (solo LiDAR and solo camera can be found in the artifact): lower perception rate means the obstacles in the round are detected in \textbf{a less reliable way} on average. We separate the perception rates for pedestrians and cars to highlight the difference. "N/A" stands for the rounds where there are no cars or no pedestrians. In general, when there are multiple obstacles presented, Apollo can "see" cars much more reliably (the median is 0.8) than pedestrians (the median is 0.3). 
For many rounds (red dots in the top figure), none of the pedestrians can be detected for the entire time. 

\begin{figure}
    \centering
    \includegraphics[width=\columnwidth,keepaspectratio]{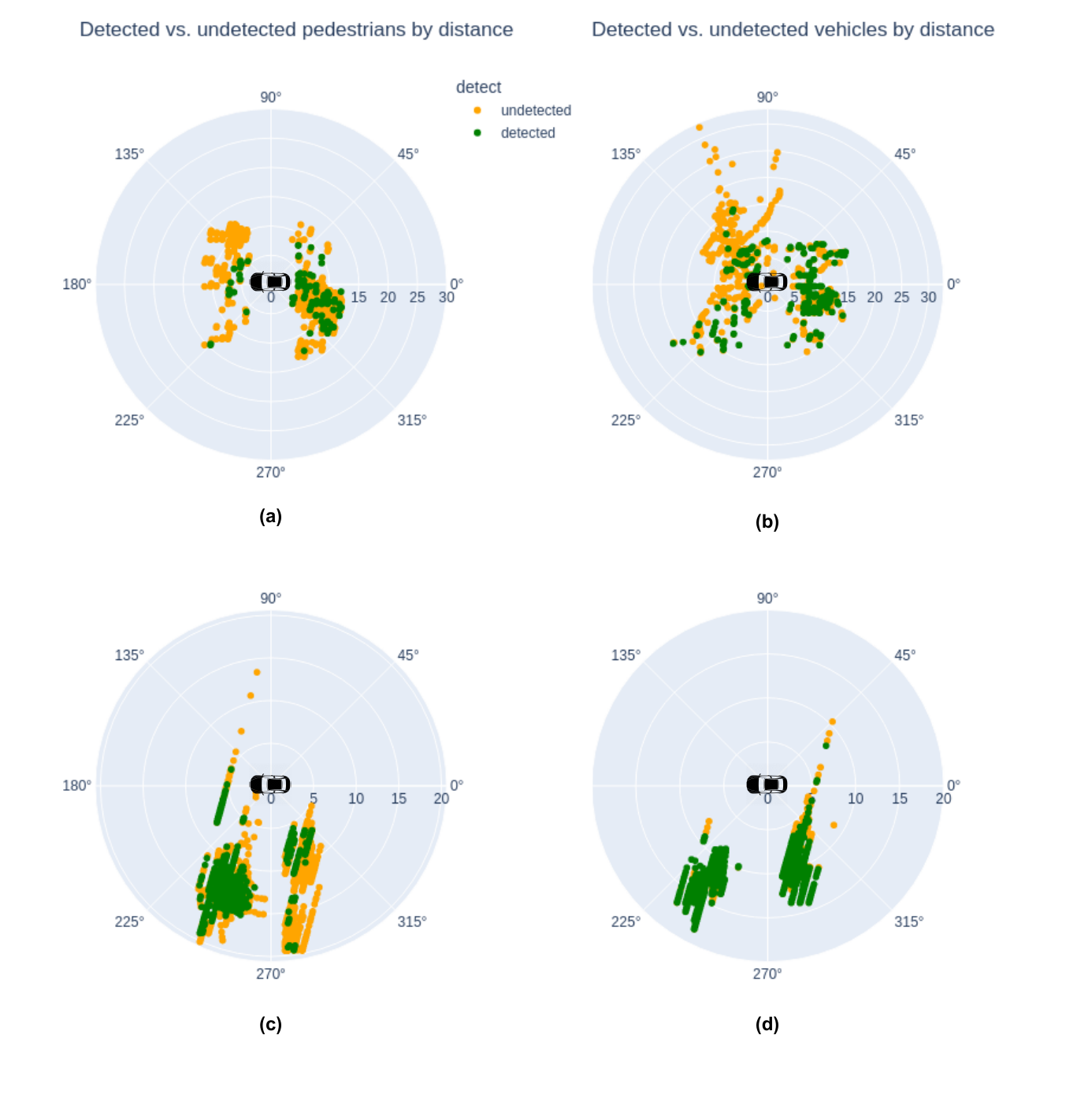}
     \caption{Graphs visualizing the distance (0-30m) between the obstacles and the ego vehicle. The black vehicle in the center represents the ego car. (a, b) are guided by maximizing avg. undetected obstacles and (c, d) are guided by maximizing neuron coverage. Pedestrians (a, c) and vehicles (b, d) are shown separately. \label{fig:neuron_distances}}
\end{figure}

Figure~\ref{fig:neuron_distances} plots all obstacles from all the frames based on their distance from the ego car. (a, b) are guided by maximizing the average undetected obstacles and (c, d) maximizing the neuron coverage. 
%For (a, b), since the fitness function is average undetected obstacles, 
We observe that (a) and (b) are very different from (c) and (d), which means that the two fitness functions guide the generation differently. Guided by neuron coverage, most of the obstacles of the generated scenes are positioned to the left side of the ego car. Differently, guided by undetected obstacles in detection range, the obstacles are more scattered to all the regions.
Such differences also indicate that the two fitness functions do not align, i.e., maximizing neuron coverage does not lead to the increase of undetected obstacles.
%similar notice the distribution of obstacles between pedestrian and vehicle across data frames is considerably different from each other. While vehicles are more dispersed, pedestrians are clustered closer together.
%Comparing this with (c, d) where the fitness function is different, we notice the distribution is more similar between vehicles and pedestrians.

\rqboxc{We find that camera-LiDAR is the best-performance perception in Apollo and the camera is the worst. Despite having been rigorously tested, \toolS finds perception weakness in Apollo. Our analysis also shows that the two fitness functions generate different scenes.}

\subsection*{RQ2: \RQThree}
\pa{Motivation.} With the perception weakness revealed by \tool, a natural question comes next as to what extent, such perception weakness may lead to adverse outcomes in simulated runs. In this RQ, we perform further analysis of the identified issues in RQ1.

%One of the most straightforward (and the most costly in terms of money and life) type of ADS malfunction is a collision either other obstacles.
%From the various fuzz guidance criteria, we can identify many seeds that has high potential to cause collision using criteria such as proximity to the ego car, velocity, whether their trajectories cross. We first attempt to identify crashes using proximity to the ego car.

\pa{Method.} We performed two further analyses on the 300 simulation rounds by \toolS in RQ1. 

First, we categorized the cases with perception weakness (i.e., there exists at least one mismatched obstacle between Apollo and ground truth) based on the distance: the closer the mismatched obstacle is to the ego car, the more serious the problem is. For example, if an obstacle is undetected by perception in a close range, the ego car would have less time to respond to avoid a collision. Or if the ego car detects a non-existent obstacle in close range, the ego car may make a hard stop.

Second, we select certain generated simulation rounds by \toolS and let them continue running for a total of 45 seconds to observe whether the simulation rounds would introduce adverse outcomes. %or until the ego car arrives at the destination. 
Our selection is based on three factors: the number of unmatched obstacles, the distance, and whether the obstacles are close enough to the trajectory of the ego car. %within a range 
Then we manually investigated the simulation rounds of adverse outcomes to identify the root cause.

\begin{figure}[htbp]
    \centering
    \includegraphics[width=0.8\columnwidth]{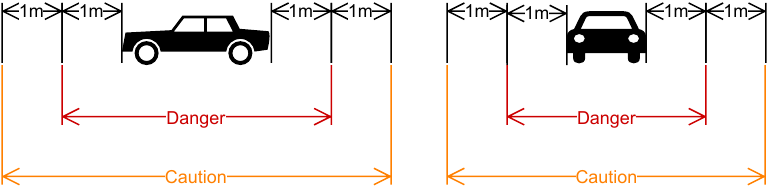}
    \caption{Different caution levels labeling for the ADS-controlled car in our framework.}
    \label{fig:caution_level}
\end{figure}

\pa{Results.} We use three categories to label the distance between an obstacle and the ego car~(Figure~\ref{fig:caution_level}): 1) "Danger": within one meter; 2) "Caution": within two meters; and 3) remaining cases. 
% \todo{we need a circular graph}

Upon investigation of the 1200 simulation rounds, we find the 11 rounds that result in three types of adverse outcomes: (1) collision, (2) unnecessary stop, and (3) wrong destination.
Below we describe our detailed analysis.

%After generating a substantial amount of seeds from the previous fuzzing rounds.
%We attempt to detect seeds that can lead to collisions.
%From all seeds, we categorize them into three types, called ``caution'' levels. There are three levels of caution that we labeled as shown in Figure \ref{fig:caution_level}:
%\begin{enumerate}
%    \item The first level is ``Danger'' level: This level is applied to a round when at any frame, there is any obstacle is within 1 meter either to the front, back, or the sides of the ego vehicle.
%    \item The second level is ``Caution'' level: This level is applied to a round when at any frame, there is any obstacle that is within 2 meters either to the front, back, or the sides of the ego vehicle.
%    \item The third level ``Ignore'' is applied to the remaining rounds.
%\end{enumerate}
\begin{figure}[htbp]
    \centering
    \includegraphics[width=\columnwidth]{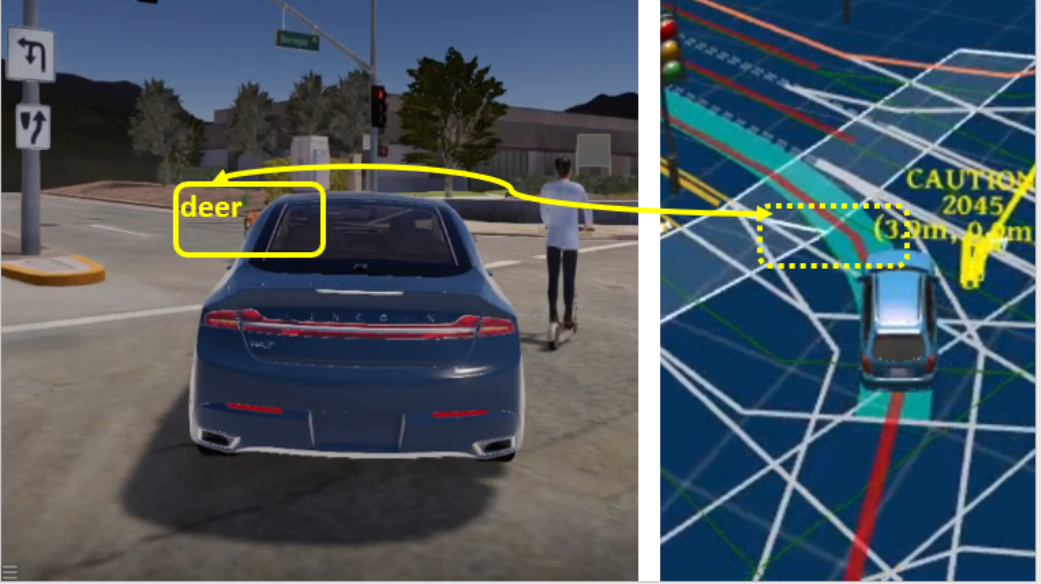}
    \caption{Apollo hit a deer since Apollo cannot see it. The left side is from the simulator (ground truth) and the right side shows Apollo detection results.}
    \label{fig:deer}
\end{figure}

\begin{figure*}[htbp]
\includegraphics[width=\textwidth]{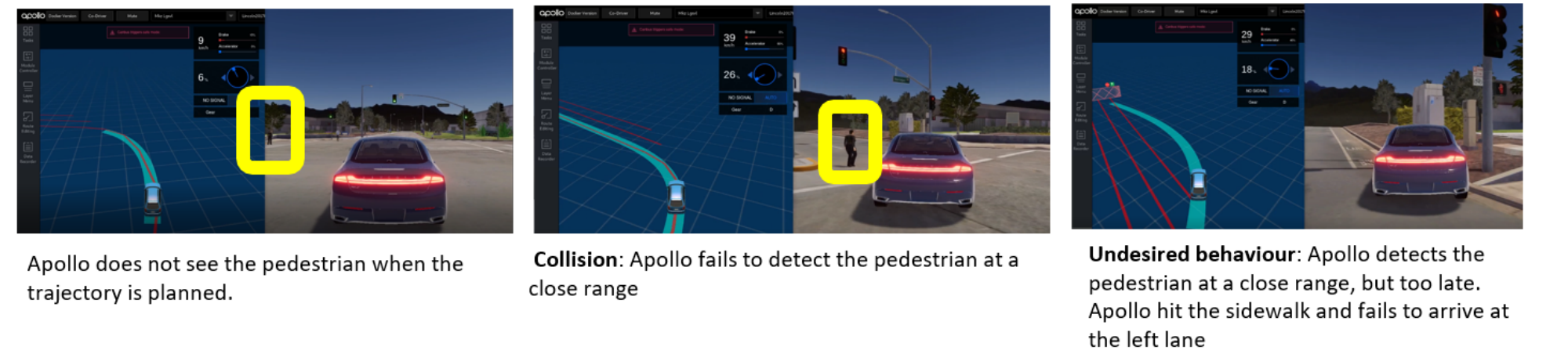}
\caption{\label{fig:collisionCloseProximity} Two examples of adverse outcomes of Apollo. The left figure shows that Apollo failed to detect the pedestrian when planning the trajectory. The right two figures show two possible adverse outcomes depending on the speed of the ego car and the distance between the pedestrian and the ego car.}
\end{figure*}

\begin{figure*}[htbp]
\includegraphics[width=\textwidth]{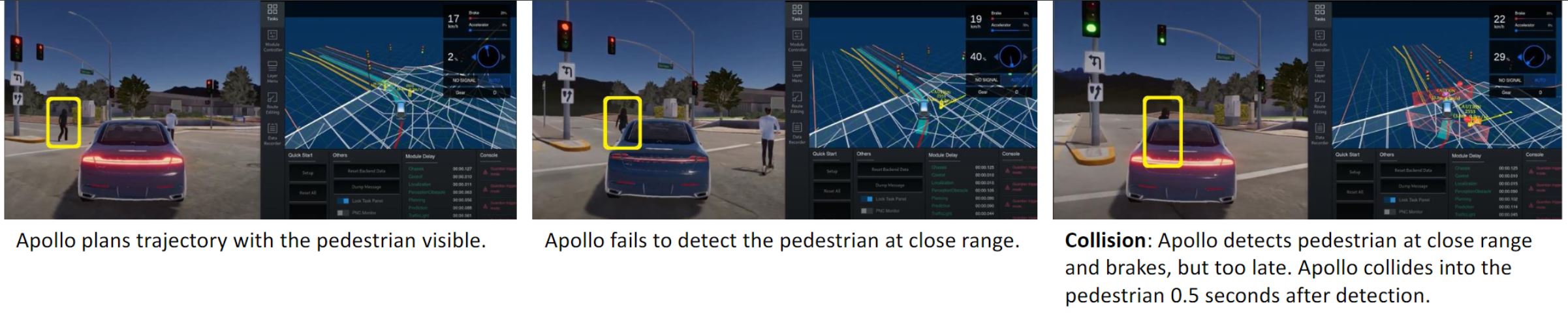}
\caption{\label{fig:round291} This example shows a collision due to inconsistent obstacle detection. The left figure shows that Apollo detects the pedestrian when planning the trajectory. The middle figure shows that Apollo's failure to detect the obstacle for a few frames caused it to plan a new trajectory without considering the pedestrian. The right figure shows that when Apollo can detect the pedestrian again, it is too late and cannot prevent the collision.}
\end{figure*}

\pa{(1)\ Collisions.}  We observed eight collision cases in the ego car hit an obstacle. We find that the reasons behind it can vary, i.e., insufficient DNN models, a combination of bad code logic and poor perception, or flaws in design logic.

\pa{- Fail to detect objects of irregular shape.} The ego car turns left and runs into a deer when the deer is crossing the street. Our investigation finds that Apollo's perception fails to detect the deer the entire time during the intersection, causing the collision. If the deer is replaced with a pedestrian of a regular height, Apollo will not end up in a collision. 
In general, we observe that Apollo's camera-LiDAR perception is not reliable in detecting obstacles of smaller size (deer and turkey modeled by \svl) compared to normal-sized pedestrians. 

%Apollo is not alone for this cause -- a recent testing video reveals that Tesla
\pa{- Trajectory too close to pedestrians.} We find that the ego car may collide with a pedestrian due to the close proximity between the planned trajectory and the pedestrian. Both the perception and planning module contribute to this issue.
Figure~\ref{fig:collisionCloseProximity} illustrates one such case. When Apollo calculates the trajectory (i.e., the blue highlighted path in the left figure), Apollo does not consider the pedestrian in black as it is not detected.
According to Apollo's planning documentation, if a pedestrian is detected when its distance from the ego car is more than 15 meters, Apollo will plan its trajectory to consider the detected pedestrian. 
Within 15 meters, Apollo would stop temporarily if it detects the pedestrian on the planned trajectory.
The center figure in Figure~\ref{fig:collisionCloseProximity} shows that the pedestrian is not detected at close range, leading to a collision. Even when Apollo can detect the pedestrian, if Apollo's trajectory is very close to the pedestrian, the situation is dangerous. 
For example, when Apollo is running at high speed, turning may cause Apollo to wobble and collide with the pedestrian in close proximity. 

\pa{- Unstable detection results.} We observe cases in which Apollo fails to detect a pedestrian \textbf{reliably} in time. In fact, Apollo has been detecting pedestrians in an unstable fashion and starts to have a reliable detection about 0.5 seconds before the collision.
However, it is too late. Figure~\ref{fig:round291} illustrates the collision case. The perception rate was 45.45\% for the 4 seconds prior to the crash, then increased to 100\% about 0.5 seconds before the crash, but not soon enough. 

\begin{figure}[htbp]
    \centering
    \includegraphics[width=\columnwidth]{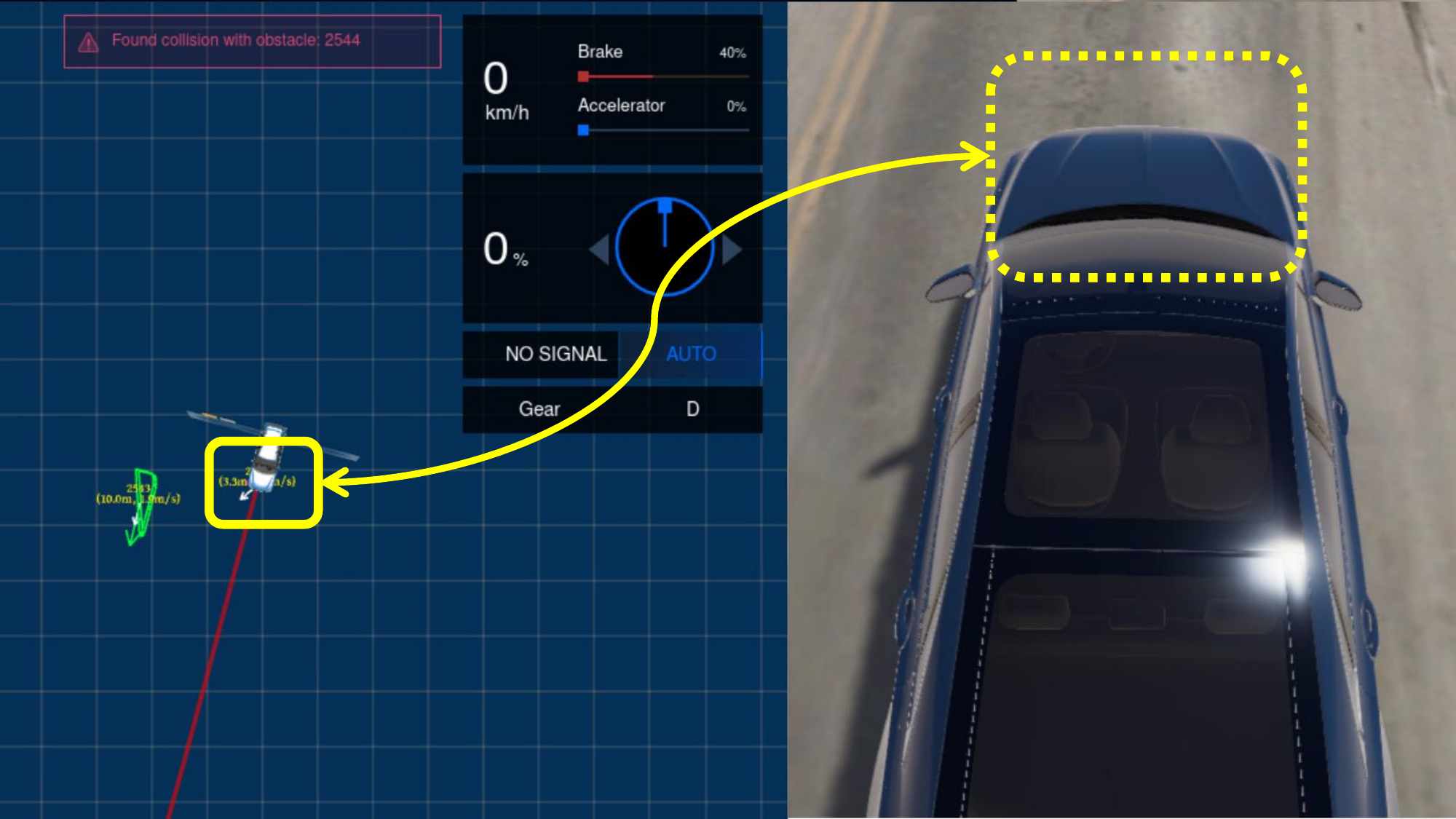}
    \caption{Example where the car stops for a non-existent obstacle.}
    \label{fig:unnecessary_stop}
\end{figure}

\pa{(2) Unnecessary stops.} We observed one case of unnecessary stops. Figure \ref{fig:unnecessary_stop} shows a non-existent obstacle (right in front of the car with id 2544), for which Apollo makes a braking decision. Unnecessary stops are less severe than collisions, but may still lead to collisions. For example, if cars fail to keep a safe distance in highway driving, an unnecessary stop would cause a rear-end collision.

\pa{(3) Undesired behavior (wrong lane).}
In four cases, we observed that Apollo failed to arrive at the predefined destination (Apollo even hit the sidewalk in one case), although that no collisions were observed. Similar to the ``trajectory too close," the cause of such cases is not single-sourced. First, Apollo only detected the pedestrian at a close range and failed to detect it further away (i.e., before deciding on trajectory). Second, although Apollo managed to slow down before hitting the pedestrian, Apollo decided to ``tail" the pedestrian while trying to follow the calculated trajectory. This causes the ego car to divert from the trajectory. In the end, depending on the speed of the ego car and other obstacles on the road, the ego car may hit the sidewalk (see the right-most figure in Figure~\ref{fig:collisionCloseProximity}) or turn into an incorrect lane.
The correct driving behavior should be waiting for the pedestrian to pass instead of ``tailing'' the pedestrian, which leads to moving to the wrong destination.

% \begin{figure}
%     \centering
%     \includegraphics[width=\columnwidth]{figures/rq/min_dist.pdf}
%     \caption{Obstacles in ground truth that are never detected by obstacle detection, grouped caution level using the closest distance possible distance of the corresponding obstacle to ego car in run.}
%     \label{fig:unmatched_groundtruth}
% \end{figure}

% \pa{(4) Obstacles that are never detected.}
% In our runs, we sometimes find obstacles that exist in the ground truth but is never detected by any obstacle detection modules. It would be useful to learn more about these obstacles and whether they can be problematic when they are undetected. Figure~\ref{fig:unmatched_groundtruth} shows such cases. In this Figure, the obstacles are grouped based on the distance as presented in Figure~\ref{fig:caution_level}. Note that the distance here is the closest that these obstacles ever get to the ego vehicle controlled by Apollo. As shown in this figure, the obstacles that are never detected by the obstacle detection modules are outside of the ``Caution'' distance, which suggests that they are at low risk to interfere with the ego car.

\rqboxc{We find three adverse outcomes of the perception weakness by \tool: collision, unnecessary stop, and fail-to-reach destination. Our manual analysis reveals that the cause is often a chain of failures instead of a single source. }
%\pa{Discussions.} Limited by time and resource (i.e., extensive manual work required for investigation), we included  
%Upon getting this data, only filter out data at the ``danger'' or ``cautious'' levels. Then, from this data, we sort the data by the average number of ground truth obstacles undetected descending take the top 10\% of these seeds, and let them continue running freely for 1 minute to see if there can be a collision given how close these obstacles were to the ego vehicle.

%\pa{Result} One case of collision was was identified is shown in Figure \ref{fig:collision1}. This case correspond to fuzzing round 297 of the random fuzzing data. There are five frames in this figure, which we will go over one-by-one.
%In the first frame, we can see that the ego car is approaching an intersection, where it detects another car in the intersection. In the second frame, as the ego car gets nearer to the obstacle, it suddenly fails to detect that obstacle. As a result, in the third frame, a collision happens. Upon closer analysis in the fourth, we can see that the LiDAR input is the same as when the ego car can still detect the obstacle, and in the fifth frame, the LiDAR input is even more intense. Yet, in both of these cases, the obstacle is not detected. Camera fails to detect this obstacle at all.

% \section{Open Science}

\section{Threats to validity}
\subsection{Internal Validity}
\pa{Seed selection and quality.}
We created a pool of original random seeds for selection and selected the seed for each round using various strategies. However, even though randomized, the pool of seed still has the potential to be skewed in certain directions, which can be exacerbated by the selection strategy.

\pa{Bugs in LGSVL.} Just like any software, LGSVL is not bug-free.
It is possible that bugs in LGSVL may cause buggy sensor input. However, note that LGSVL is relatively well tested and mature, and also very commonly used in prior work on testing ADS~\cite{avfuzz, adfuzz, metamorphic_lidar, metamorphic_prediction}.
In addition, for the issues we detected on Apollo, we manually verified that such cases were valid and not impacted by bugs or delays in LGSVL. In particular, we logged the sensor data in multiple locations, i.e., when generated by LGSVL, when being passed to Apollo, and when received by perception, and verified such sensor data is consistent in all such locations. Also, we verified such sensor data does match with the ground-truth obstacles. 
%of such and find that the obstacle detection modules receive near identical data in such scenes and the obstacles do not disappear/reappear in the scene.

% \pa{Time consumption.}
% \pa{Synchronization}

% \vspace{-0.1in}
\subsection{External Validity}
% While the simulator is the closest we can get to real world testing, as evident by our usage.

\pa{Applying \toolS on other ADS and simulators.} \toolS can be integrated with other simulators such as CARLA~\cite{carla} as CARLA have the same functionalities as LGSVL.
%From their documentation, CARLA~\cite{carla} shows that they have the same functionality as LGSVL that we use for our experiment. Combine this with its ability to work with Apollo~\cite{caral_apollo_bridge} and openpilot, we believe our approach can be replicated on CARLA~\cite{carla}. However, we do not have this information on other simulators.
\toolS can also be extended to work with LGSVL-Autoware and CARLA-openpilot. Our experiment is limited to LGSVL-Apollo. It remains as future work to experiment on such combinations.%However, we do not have this information on other ADS. \todo{polish}

%\pa{Hardware.} While our system is sufficient to run Apollo, there are delays in the messages our framework received due to 
%the complex operations in Apollo.
%We address this by only matching messages within a tolerable time difference (30 ms).

\pa{Open Source ADS.} Our experiment is limited to one open-source ADS Apollo. While the exact findings may not generalize to other proprietary systems, we believe \toolS can be applied to multi-module proprietary ADS as simulation is the defacto testing approach in the ADS industry. Moreover, bugs found in open-source ADS can indicate underlying issues in proprietary systems as shown in \cite{metamorphic_lidar} where it reveals that obstacle detection models in ADS may not have been adequately trained with noisy inputs.

% \pa{Obstacles Detected by Sensors and the Ground Truth.}
% The ground truth is acquired from the simulator which is the list of obstacles it added to the scene. It is possible the 

%\subsection{Construction Validity}

%\pa{Matching ground truth to perception results.} 
%When comparing Apollo's obstacle detection with ground truth, in each pair of frames, we match the obstacles detected in perception to their counter parts in the ground truth using Hungarian algorithm, which is also used by Apollo's tracking module. 
%It is possible that Hungarian algorithm  produces incorrect matches. We used a very small distance in Hungarian Algorithm to reduce the possibility of incorrect matches and manually verified the cases found in RQ2.

%\pa{Grey-Box Testing.}

%\pa{Simulation Speed.}
%As described earlier, when there is need to compute neuron coverage, we slow down the time in the simulation. While we do not see evidence of it impacting the performance of Apollo in our experiment, we are not sure about cases not observed.

% \vspace{-0.03in}
\section{Discussions and Future Work}%\todo{delete if not enough spsace}
%\pa{Needs for debugging and diagnosis support.}

%\pa{Needs for better documentation and requirements.}

\pa{The naturalness of the generated scenes by \tool.} We add a few constraints to make the generated scenes realistic. First, we control the number of obstacles in each scene not exceeding 15. Second, when performing mutation operators involving modifying the location of the obstacle, we make sure no obstacles will overlap. Also, the simulator (LGSVL) makes sure all the obstacles (pedestrians/vehicles) follow traffic rules and have valid destinations. In the future, we plan to extract real-world driving scenes from recordings~\cite{kitti} to further diversify and improve the naturalness of the generated scenes. It is also interesting to explore how to decide whether a generated scene is natural or not, quantitatively or following guidelines. 

%\pa{Testing a highly configurable ADS.}

\pa{Reality v.s. high-fidelity simulator.} Modern simulators are high-fidelity thanks to the advanced game engine and powerful computing resources. Still, there exists a gap between reality and simulators. It is well known that LiDAR sensor data is highly impacted by surface materials, lighting, and weather conditions. For example, rain would introduce non-trivial perturbations to LiDAR sensor data. Future work should investigate the gap between LiDAR in reality and LiDAR in simulators. A similar study is performed to compare the camera images between reality and simulators~\cite{gap}. 
%Hence, the input to Apollo's perception is constrained by how close the simulation is to the real world, the ability of the simulated sensors to capture the simulated scenes, and the level of agreement between the sensors. 
%\jinqiu{This reads vague.This paragraph reads a mix of internal and external validity.}

\pa{Neuron coverage.} It is more difficult for \toolS to achieve a more diverse set of neurons, i.e., activating specific neurons, compared to white-box techniques~\cite{deepxplore}. For example, through 300 generated scenes, the neuron coverage of the LiDAR model of Apollo improves from 81\% to 83\%.
It is possible to further improve neuron coverage by adding minor perturbations to the sensor data. However, we view neuron coverage as a way to guide test generation, i.e., generating more unique test scenarios, as a recent study~\cite{neuron_coverage} shows that further improving neuron coverage may not necessarily improve bug detection capability. 
%since we are testing the perception module using scenes with objects instead of transforming or adding noise to existing sensor data, this is expected. Also, perturbations or transformations of the input is not particularly meaningful for our tests.

%\begin{figure}[htbp]
%\includegraphics[width=\columnwidth]{figures/rq/coverage.pdf}
%\caption{Comparison of Neuron Coverage between non-guided \toolS (random) and guided \toolS (neuron). Both are for camera-LiDAR fusion.\label{fig:coverage_comparison}}
%\end{figure}
%First, we compare the neuron coverage between the random non-guided \toolS and neuron-coverage guided \toolS. Figure~\ref{fig:coverage_comparison} shows the neuron coverage between the two. Both are based on camera-LiDAR fusion, which means both of the models (LiDAR and camera) are used for neuron coverage calculation.
%We can see that in the first few rounds, neuron coverage from guided \toolS does not increase any faster than the non-guided \tool, probably due to the fact that non-guided \tool has a better neuron coverage from the beginning. However, over time, neuron coverage of guided \toolS improves faster than the non-guided one.
% \vspace{-0.05in}
\section{Conclusions}
In this paper, we present a fuzzing technique (\tool) to test a multi-module ADS. \toolS focuses on finding weakness in the perception module and further exercises how such weakness will impact the subsequent modules of an ADS and eventually the safety of ADS. We applied \toolS on an industry-grade L4 ADS (Baidu's Apollo).
Leveraging a priority strategy to rank the generated hundreds of test runs, we reproduced test runs that exposed critical even fatal failures in Apollo. 
%These errors can cause faulty behavior or collisions. 
We diagnosed the detected issues and discussed how Apollo can fix such issues.%For the problems identified, we can localize the cause of the issue to the related sensor or fusion. This data can help developers in quickly fixing these issues without having to spend significant amount of time locating the bugs.

\section*{Data Availability}
 We release the following data from our work:~\footnote{Artifact from this work is released at https://github.com/myproxemail2022/sim-test-ads}
 \begin{itemize}
     \item Generated simulation runs by \tool
     \item Data (coverage, precision, recall perception rate, etc.) of simulation runs
     \item Recordings and analysis on the found problems of Apollo
 \end{itemize}

% \balance

\bibliographystyle{ACM-Reference-Format}
% \bibliography{paper}

\end{document}